\documentclass[showpacs,aps,prb,twocolumn,groupedaddress]{revtex4-2}
\usepackage{graphicx}
\usepackage{textcomp}

\begin{document}


\title{Morphology and Magnetic vortex chiral symmetry of 2D arrays of magnetic trilayer disks with magnetostatic interlayer coupling determined by  X ray resonant magnetic scattering }


\email[]{jidiaz@uniovi.es}
\author{J. D\'iaz$^{1,2}$, L. M. \'Alvarez-Prado$^{1,2}$, S. M. Valvidares$^{3}$, I. Montoya$^{4}$, C. Redondo$^{4}$, R. Morales$^{5,6}$ and M. V\'elez$^{1,2}$}
\affiliation{$^{1}$Universidad de Oviedo, Calle Federico Garc\'ia Lorca 18, 33007, Oviedo}
\affiliation{$^{2}$CINN (CSIC – Universidad de Oviedo), 33940 El Entrego, Spain}
\affiliation{$^{3}$ALBA Synchrotron, 08290 Cerdanyola del Vall\'es, Spain}
\affiliation{$^{4}$Dept. of Physical Chemistry, Univ. of the Basque Country UPV/EHU, E-48940 Leioa, Spain}
\affiliation{$^{5}$Dept. of Physical Chemistry, Univ. of the Basque Country UPV/EHU and BCMaterials, E-48940 Leioa, Spain.}
\affiliation{$^{6}$IKERBASQUE, Basque Foundation for Science, E-48011 Bilbao, Spain}


\date{\today}

\begin{abstract}
X ray resonant magnetic scattering (XRMS) was used to characterize the magnetization of 2D arrays of trilayer submicron magnets. The interpretation of the data required the understanding of the morphology of the magnets which was also deduced from the scattered intensity. The magnets consisted of two magnetostatically coupled ferromagnetic layers separated by a non-magnetic spacer. The scattered intensity from the disks resulted to be dependent on the disks surface curvature. This made the collected intensity at each Bragg reflection (BR) to be correlated to the reflected light from locations of the disk with the same angle of curvature. Due to this, quantitative information was obtained, averaged over the disks illuminated by x rays, of the variations in thickness and magnetization across the entire area of the disks. This averaged magnetization mapping of the disks served to study their vortex configuration in each of their magnetic layers, determining the average location of the vortex, the chiral symmetry of its magnetic circulation, and the specific locations where the vortex nucleation starts within the disks. Chiral asymmetry appeared in the disks when the field was oriented at an oblique angle with respect to the easy axis of the array. The local magnetic sensitivity of the technique allowed to identify a non-centrosymmetric distribution of the magnetization of the disks that explains the observed chiral asymmetry. Unexpectedly, the magnetic circulation sense of the vortex was the same in both ferromagnetic layers. In addition, the magnetization of the buried layer was different in the descent branch than in the ascent branch of its hysteresis loops. This effect was also found in some of the hysteresis loops of both layers collected at different BRs in two different sample orientations, suggesting that the magnetization and demagnetization of the disks could be affected by collective stochastic process. 
\end{abstract}


\maketitle

\section{Introduction}

Magnetic vortices formed in simple magnet forms have been the subject of investigation since the methods to create arrays of submicron size magnets are available \cite{2001_Field_evolution,2001_Magnetic_switching,2002_Vortex_chirality,2005_Magnetic_remanent,2008_Magnetic_Vortex}. They are a stable magnetic configuration in disks and squares magnets, and they are relatively simply to describe. An interesting aspect of the vortex is that it can adopt 4 different configurations, attending to the handiness of the magnetic rotation and the sense of polarization of its core. These configurations are robust and they have been intended to be used for information storage \cite{2000_Ultrahigh_density,2010_A_frequency_controlled}, spin wave source \cite{2005_Dynamics_of_coupled,2016_Magnetic_vortex_cores}, magnetic sensors \cite{2018_Topologically_protected,2022_Control_of_sensitivity,2019_Evolution}, and  biotechnology \cite{2009_Biofunctionalized_magnetic_vortex,2020_Magnetic_nanostructures}. The difference in energy between the configurations depends on the symmetries of the system in many cases, and its characterization and control has been an important subject of investigation for years. 

The behavior of stacking magnetic layers in the same shaped magnet is, however, less known. A double layer system allows more parameters to tune, like, for instance, the interaction between layers through the non-magnetic spacer and the magnetization of the disks, increasing its functional capabilities \cite{2010_Control_of_Double_Vortex,2022_Control_of_sensitivity}. Such a double layer structure is actually the chosen in sensors and memory units like in spin valves and magnetic tunnel junctions. However, the studies where more than one magnetic layer in the disks are involved are scarce due to the more difficult characterization of the buried layers \cite{2005_Dynamics_of_coupled,2008_Magnetostatic_and_exchange, 2022_Unusual_Magnetic_Hysteresis_and_Transition_between_Vortex_and_Double_Pole_States}.  Most of the magnetic sensitive microscopies with submicron resolution are surface sensitive, like MFM (Magentic Force Microscopy) \cite{2019_Frontiers,2020_Probing_the_pinning} and PEEM (Photoemission Electron Microscopy) \cite{2010_Switching_a_magnetic_vortex}. Transmission electron microscopies are not layer sensitive  \cite{2000_Lorentz_microscopy,2001_Magnetic_switching,2010_Control_of_Double_Vortex}. They are also subjected to limitations in the substrates, which have to be transparent to electrons, and the applied fields during measurements. These limitations are overcome in XMRS, making it the tool of choice for the characterization of these systems due to its capability to peer into buried layers at relatively large thickness. The present experiment also shows that XRMS can have enough lateral resolution to register local changes in the magnetization of submicron size magnets, averaged over all the probed magnets.

XMRS is a non-destructive photon-in photon-out technique, which makes it compatible with the use of external fields, currents or temperature during measurements. The only restriction for the sample substrates is to be flat. The principles of magnetic scattering are similar to the observed using light in the visible spectrum \cite{1993_TMOKE_hysteresis,2002_Pure_magneto_optic,2004_The_diffracted_magnetoptic_Kerr,2007_Vector_and_Bragg_MOKE,2008_Extended_longitudinal}.  Changing the polarization of the incident beam allows either to measure the longitudinal or the transverse component of the magnetization of the probed magnets \cite{2007_Vector_and_Bragg_MOKE,2008_Soft_X_ray_resonant_magnetic}. Using x rays permits the access to a wider range of moment transfer values, increasing the sensitivity to local changes. To distinguish the magnetic signal from each of the stacking layers, the energy and polarization of the x rays must be tuned, requiring a synchrotron radiation source. XRMS has been already used in this kind of systems before \cite{2006_Layer_resolved,2007_The_breakdown, 2008_Magnetostatic_and_exchange}. In particular, the study done in reference \cite{2007_X_ray_resonant_magnetic_scattering} is the only one that recovers the magnetization of each layer in a bilayer square ring magnet at different subregions by deconvoluting the contribution of each of this subregions to the intensity of the diffracted spots. This method requires magnets with geometrically  well differentiated regions and form factors. In all the cases, it is assumed that the magnets are flat, i.e., perfectly bidimensional. However, this is not always the case. Short wave length sources can be specially sensitive to this. The present study deepens in the interpretation of the intensity obtained at different x ray Bragg reflected angles for those cases in which the form of the magnets is not perfectly flat, finding a correlation between the angle at which the Bragg reflection (BR) is collected and the region of the submicron magnet from where the light comes. For those case, this converts XRMS in a magnetic microscope capable to see the variations in the lateral component of the magnetization, averaged over all the submicron magnets illuminated by the x rays, at specific regions of the submicron magnets and at specific layers. It also evidenced the sensitivity of XRMS to the morphology of the disks in a quantitative way, making possible to determine changes in the thickness across the area of the submicron magnets with nanometer resolution. This is an aspect that it has been largely overlooked, since most of the studies done assumed flat surfaces in their magnetic forms. 


A direct consequence of this finding has to do with the sensitivity of the XRMS to the chiral asymmetry of the vortex. This was demonstrated by us in a previous experiment on an array of a single layer of Permalloy disks \cite{chiral_nanotech}. This sensitivity arises in XRMS from the origin of the magnetic scattering intensity in these systems, which is due to the interference between the magnetic and charge scattering, and it has a similar origin than the observed using visible light \cite{2007_Vector_and_Bragg_MOKE,2008_Extended_longitudinal}. Although this interpretation is still perfectly valid in flat or nearly flat forms, the present experiment shows that chiral sensitivity is enhanced by the curved surface of the submicron magnets. 

In the studied sample, chiral asymmetry was detected when the disks array was oriented at an oblique angle with respect to its easy axis, changing its vortex chiral sense with the direction of the initial magnetization in saturation. Thanks to the here reported microscopic sensitivity of XRMS, the location of the nucleation area in the disks was determined, demonstrating that the magnetic chiral asymmetry of the disks is associated to a non-centrosymetric distribution of their magnetization. Chiral symmetry was also detected in each of the layers, resulting to be the same in both of them, what was unexpected specially due to the nature of the observed chiral asymmetry. The measured hysteresis loop branches in some locations of the disk, or even in all the disk, were not symmetric, suggesting a collective stochastic behavior in the magnetization and demagnetization of the disks, probably induced by thermal fluctuations during measurements\cite{2007_The_breakdown}.

\section{Sample preparation and characterization}
The array of disks was produced by, first, creating an antidot array by interference laser lithography (ILL) on a negative resist that was spin wet on silicon substrates. The ILL procedure used a Lloyd's mirror interferometer with a He-Cd laser ($\lambda$ =325 nm) as the light source \cite{2018_Cost_Effective}. ILL produces patterns of a constant period and similar disks shape over large areas of the order of $cm^{2}$ in a single shot. In this way, the x ray beam had not restrictions in its size to probe the samples. The metallic layers were deposited on these antidot imprinted substrates by  magnetron sputtering. The measured disks array resulted after the resist was lift off. Evaporation of each layer was done at normal incidence in a vacuum chamber at a base pressure of $1\times 10^{-7}$ mbar and under an Ar pressure of $3\times 10^{-3}$ mbar. The iron layer was 14.8 nm thick, and it was deposited directly on the substrate with no buffer layer, following by the deposited of the aluminum spacer layer (2.2 nm) and the cobalt layer (17.6 nm). A 3 nm aluminum capping layer was deposited on top to avoid contamination. 

\begin{figure}
\includegraphics[width=8 cm]{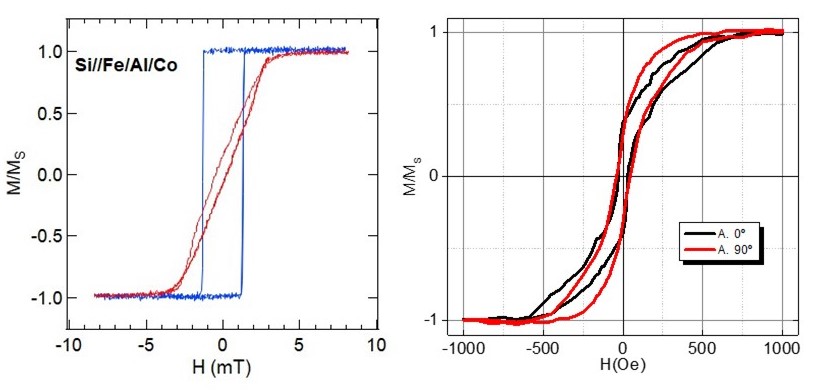}
\caption{ (a) Hysteresis loops of a  reference thin film prepared at the same time that the array of disks; (b) VSM hysteresis loops of the 2D array of disks  along the (10) and (01) direction of the square lattice. \label{Kerr_VSM_2Ddisks}}
\end{figure}

A reference sample was deposited at the same time than the substrates with the imprinted pattern. Figure \ref{Kerr_VSM_2Ddisks} shows the hysteresis loops of the reference and the patterned samples measured by VSM. The reference sample was magnetically soft, with an in-plane magnetic anisotropy of about 40 Oe and a coercive field in the easy axis (EA) of about 15 Oe. The hysteresis loops of the array of magnetic disks are the expected in magnetic configurations that minimize their stray magnetic fields, with low remanence, coercivity and comparatively large saturation fields. Its relative remanence $M_{r}/M$ is of the order of 30$\%$, and its coercive field is of about 20 Oe. The saturation field of the sample changes from 550 Oe to more than 700 Oe at two orthogonal directions parallel to the symmetry axis of the square lattice of the array. 

\begin{figure}
\includegraphics[width=8 cm]{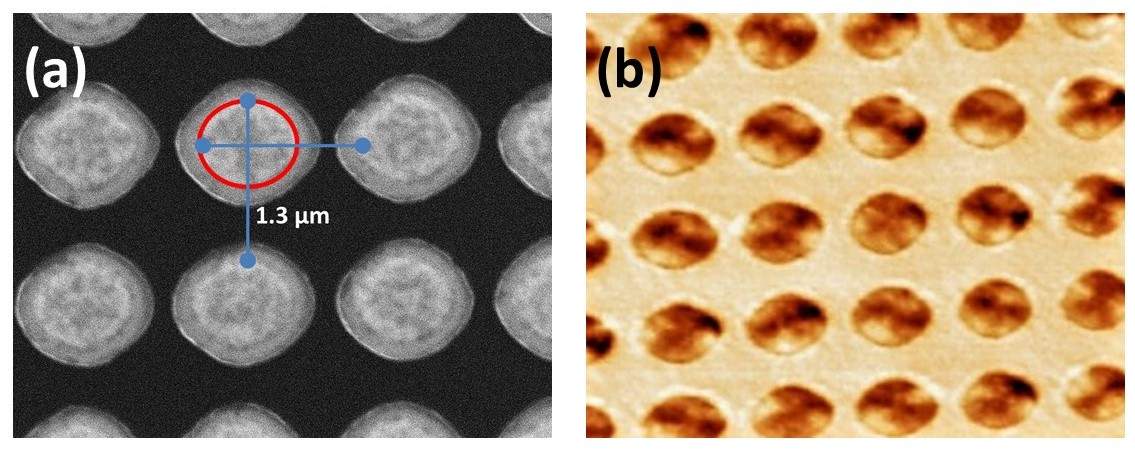}
\caption{ (a) SEM image of the array of disks; (b) MFM image of the array of disks. \label{SEM_MFM_2Ddisks}}
\end{figure}

Scanning electron microspe (SEM) images show that the lattice is perfectly squared with a lattice parameter, $\alpha$, of 1.3 $\mu m$, confirmed by the x ray diffracted pattern. The magnets have a diamond shape with rounded corners. The corners are aligned to the square lattice axis. The axis of the magnet related to these directions of the array have not the same length: their proportion ratio  is of about 0.9 (see figure \ref{SEM_MFM_2Ddisks}). The size of the disks was of 805 nm in the long axis. The EA of the sample was parallel to this axis. MFM images confirmed the presence of a single vortex at the top layer of most of the disks in the remanent state of the samples  (see figure \ref{SEM_MFM_2Ddisks}). 

\begin{figure}
\includegraphics[width=8 cm]{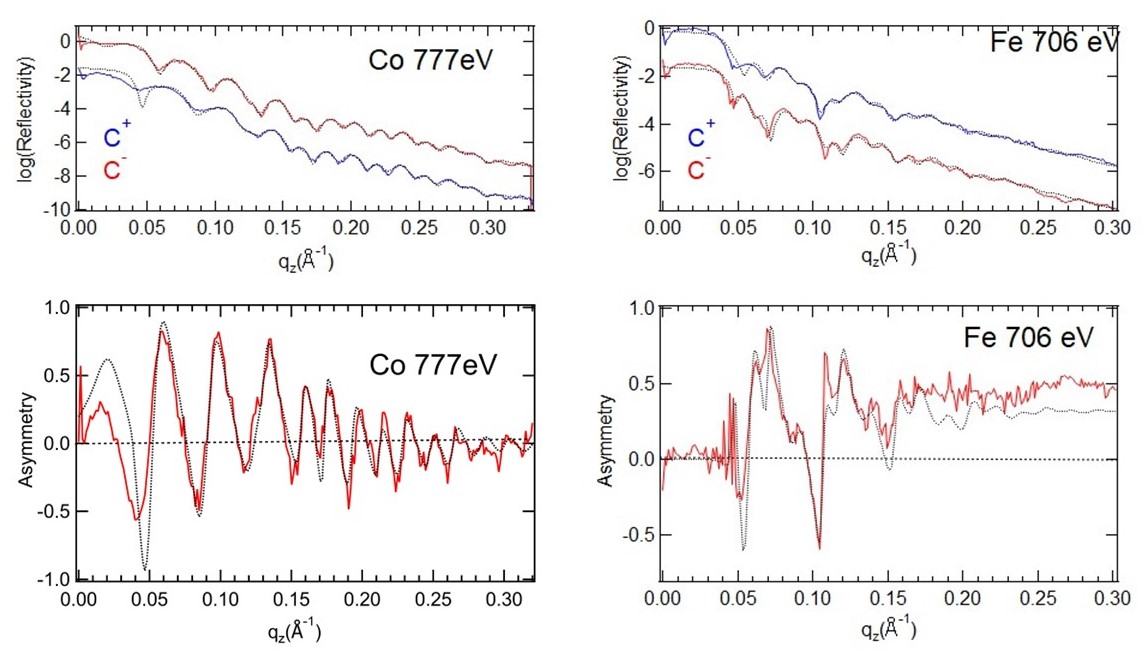}
\caption{ Fitted reflectivity and magnetic asymmetry curves of the reference thin film taken at the (a) Co (776 eV) and (b) Fe (706 eV) edges . \label{rflx_fit_thinfilm}}
\end{figure}

The thickness of the deposited layer and the quality of their interfaces was obtained by fitting the reflectivity curves of the reference sample taken at the Fe and Co resonant energies using circularly polarized light. Figure \ref{rflx_fit_thinfilm} displays the reflectivity curves and the magnetic dichroism asymmetry with their corresponding fitting curves obtained at the two resonant energies of Fe (706 eV) and Co (777 eV). The fitting of the curves was done by a home-made code using the methodology presented in reference \cite{2011_X_ray_resonant_magnetic}. The thickness of the cobalt and iron layers were 176 \AA\  and 148 \AA\  respectively. Their thickness ratio was chosen to be approximately the same that their magnetization ratio, which is 0.84, to have the same magnetization in the two layers. The thickness of the aluminum spacer was of about 22 \AA, which was large enough to consider that the magnetic interaction between layers was entirely dipolar. The interfaces of the cobalt and Fe layers with the nonmagnetic aluminum spacer had a roughness of about 9 \AA. This value might include some possible intermixing between layers. 

\begin{figure}
\includegraphics[width=8 cm]{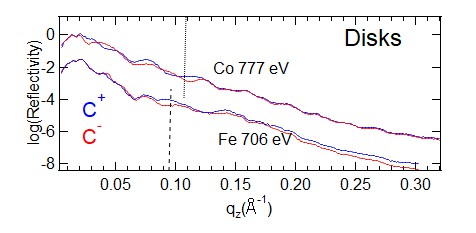}
\caption{Reflectivity curves of the array of disks taken at the cobalt (776 eV) and iron (706 eV) edges. The position in $q_{z} (\theta_{i})$ chosen for obtained the diffraction patterns with magnetic contrast are marked in each curve with a vertical line.  \label{XReflex_Disks_FeyCo}}
\end{figure}

Resonant magnetic reflectivity curves were also obtained from the disks, which are displayed in figure \ref{XReflex_Disks_FeyCo}. Oscillations due to magnetic contrast were visible and they were used to determine the x ray incident angle under which magnetic contrast was the highest in the range of large $q_{x}$ values. However, the curves cannot be fitted in the same way than the reference sample, likely due to the discrete nature of the probed surface which might introduce intensity unrelated to pure reflection from the disks. 

\section{Magnetic Scattering}
The scattering of x rays is sensitive to the magnetic moment of the probed material by considering the terms of the photon scattering that are sensitive to the angular moment of the electrons. These terms are usually too small but they are enhanced when the photon energy is close to the absorption edge of the probed element where its magnetic properties are better manifested, the scattering becoming element specific. In this case, the chosen photon energies were 706 eV  and 777 eV, the energies of the $L_{3}$ edge for Fe and Co, respectively. Using circular polarized light and grazing incidence, the intensity observed is, to a good approximation, proportional to the scalar product between the scattered magnetic moment of the electron, $\vec{m}_{3d}$, and the wave vector of the incident beam, $\vec{k}_{i}$. In the present experiment, magnetic contrast was obtained at the BR directions only, indicating that this king of scattering corresponds to the  term related to the interference between the charge scattering and the magnetic scattering. This scattered intensity, $I_{MQi}$, has the following dependence on the scattering vectors and the magnetic configuration of the disks (\cite{2008_Soft_X_ray_resonant_magnetic}):

\begin{widetext}
\begin{equation}
I_{MQi}=Re\left[F_{0}^{*}\rho^{*}\left(\vec{q}\right) F_{1}\left(\vec{k}_{i}\cdot \textbf{M}\left(\vec{q}\right)+\vec{k}_{sct}\cdot\textbf{M}\left(\vec{q}\right)\left(\vec{k}_{i}\cdot\vec{k}_{sct} \right) \right) \right] P_{3}\approx 2Re\left[F_{0}^{*}\rho^{*}\left( \vec{q} \right) F_{1}\vec{k}_{i}\cdot\textbf{M} \left(\vec{q}\right)  \right] P_{3}
\label{equation01}   
\end{equation}
\end{widetext}

$\textbf{M}\left(\vec{q}\right)$ and $\rho^{*}\left(\vec{q}\right)$ are the Fourier transform of the magnetic and charge configuration of the array of disks, $F_{0}$ and $F_{1}$ are the scattering factors for charge and magnetic scattering, $\vec{k}_{i}$ and $\vec{k}_{sct}$ are  the wavevectors of the incident and the scattered beams, respectively. $P_{3}$ is the circular polarization degree and $\vec{q}=\vec{k}_{sct}-\vec{k}_{i}$ is the moment transfer vector. 

The structure and magnetic characterization of the samples was done measuring the reflectivity curves over a relatively wide range of $2\theta$ angles, from $0^{\circ}$ to $50^{\circ}$using a photodiode detector. The light scattered from the two dimensional array was detected using a CCD camera placed at the same location than the photodiode. Magnetic fields were set at constant values for each measurement using a dedicated electromagnet \cite{BOREAS_setup}. Magnetic contrast at the BRs in the CCD images at each magnetic field intensity was obtained by calculating the circular dichroism: $I_{M}(\vec{q})=I^{C^{+}}_{MQi}-I^{C^{-}}_{MQi}$. Charge scattering was extracted summing the intensity obtained at the two circular polarization helicities, $I_{M}(\vec{q})=I^{C^{+}}_{MQi}+I^{C^{-}}_{MQi}$. 

\begin{figure}
\includegraphics[width=8 cm]{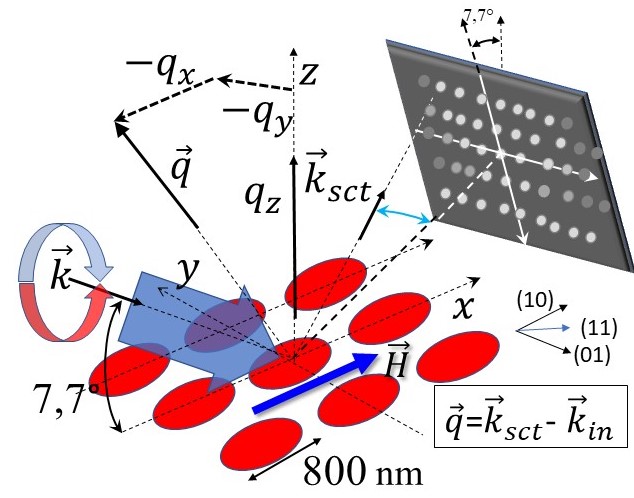}
\caption{ Geometry of the scattering experiment: orientation of the $q_{x}$ and $q_{y}$ axis with respect to the incident, $\vec{k}_{i}$, and scattered, $\vec{k}_{sct}$, beams.  \label{geometry}}
\end{figure}

The photon energy and incidence angle chosen were those in which the dichroism contrast occurred at a similar angle for the Co and Fe edges to insure that the probed areas of the sample were nearly the same. The chosen angle of incidence was  $\theta_{i}=7.7^{\circ}$, which was high enough to include a large number of BRs in the $q_{x}$ direction, parallel to the direction of the beam. Figure \ref{geometry} shows the geometry and the axis orientation for the $q_{x}$ and $q_{y}$ moment transfer vectors used in the CCD images. 

Figure \ref{XReflex_Disks_FeyCo} indicates, in the $q_{z}$ scale ($q_{z}=2k_{0}\sin \theta_{i}, k_{0}=2\pi/\lambda$), the dichroism contrast for that angle of incidence at the iron and cobalt chosen photon energies. A total of 10 snapshots with 0.1' exposure were recorded to obtain the final Bragg intensity pattern covered by the CCD camera at each of the applied magnetic fields for each circular polarization helicity, $C^{+}$ and $C^{-}$, of the incident x-rays. This process was repeated at different applied fields to complete the two branches of an hysteresis loop (HL). Each of the branches consisted of 30 measurements at constant increments of the applied magnetic field, where $\Delta H=$67 Oe. HL started measuring at magnetic saturation fields of 1 kOe. The measurements were done in two different orientations of the disks arrays with respect to the magnetic field: 1) at the (11) orientation of the array, oblique to its magnetic easy axis (EA), and 2) at the (10) orientation, parallel to the EA.

\section{Bragg Reflexions}

\begin{figure}
\includegraphics[width=8 cm]{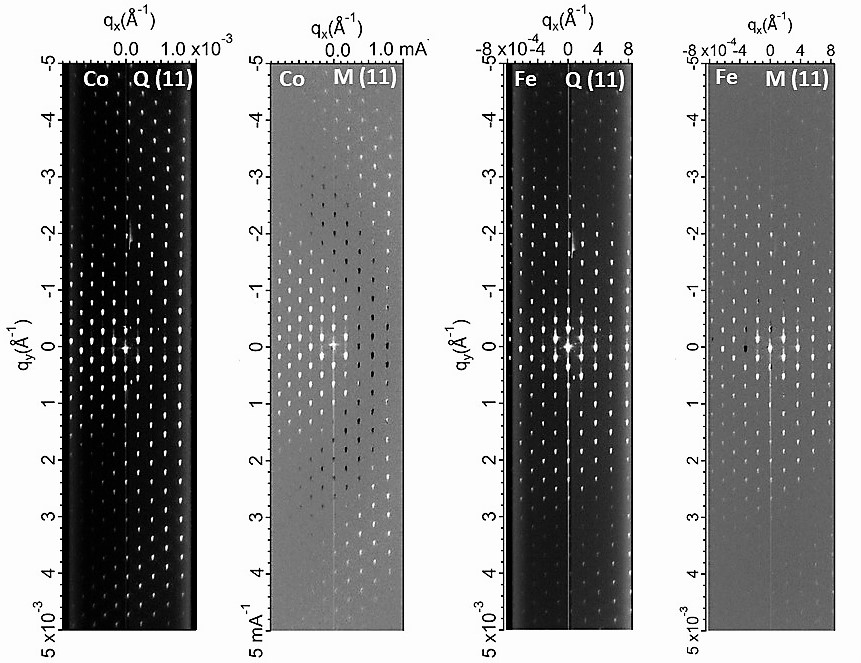}
\caption{ Diffraction pattern of the 2D array of disks taken at the Co (776 eV) and Fe (706 eV) edges at the (11) orientations due to charge scattering (Q) and magnetic scattering (M). The sample was magnetically saturated during image acquisition. \label{CCD_QyM_CoyFe11}}
\end{figure}

\begin{figure}
\includegraphics[width=8 cm]{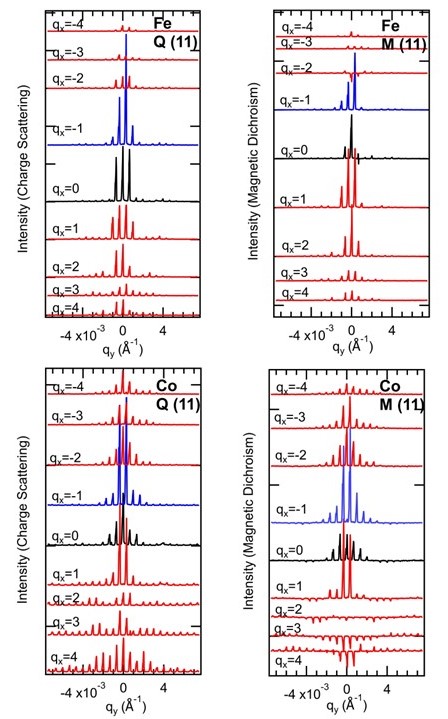}
\caption{Charge (Q) and magnetic dichroism (M) scans along the $q_{y}$ direction for different values of $q_{x}$ in iron and cobalt in the (11) orientation \label{Perfiles_FeyCo11}}
\end{figure}

Figure \ref{CCD_QyM_CoyFe11} and \ref{Perfiles_FeyCo11} display the diffraction patterns recorded on the CCD at the Co and Fe resonant energies, 776 eV and 706 eV, respectively in the (11) array orientations with their corresponding profiles along the $q_{y}$ direction for all the $q_{x}$ values collected in Fe and Co at this orientation.  The figures contain the diffraction pattern due to the scattering of the charge ($I_{Q}\left(\vec{q}\right)$) and the magnetic dichroism ($I_{M}\left(\vec{q}\right)$) obtained with the sample in magnetic saturation. The images have been scaled in the $q_{x}$ and $q_{y}$ components of the moment transfer. Due to the grazing incidence geometry, the range of $q_{x}$ values collected by the CCD camera is smaller than in the $q_{y}$ direction. The dynamic range in the CCD images have been reduced to enhance the intensity of Bragg reflections at angles far from the reflected beam, which is the most intense. 

The intensity of the diffraction pattern spreads from the center of the image, with more intensity at positive than at negative $q_{x}$ values. This intensity is also modulated, i.e., its value oscillates from the center of the pattern, apparently forming parabolic curves with their vertex located at the positive side of the $q_{x}$ axis. Both (10) and (11) orientations have a similar intensity distribution pattern. There exists marked differences between the distributed intensity in Co and Fe: the intensity at the center looks broader in cobalt than in iron when moving to negative $q_{x}$ values. In Fe, the BRs with the lowest intensity near the center of the diffraction pattern forms an incomplete ring. Cobalt magnetic contrast changes sign twice counting from the center to the border. Fe changes the sign of its magnetic contrast in few points just at the center of the image, but there is no change in contrast at higher $q_{y}$ values as in cobalt. 

The observed modulation in intensity of the BR peaks cannot be attributed to the form factor of the disks. Actually, such a diffraction is missing from the pattern. If it existed, it should form concentric rings from the center of the pattern since its form factor depends only on the in-plane coordinates $x$ and $y$, i.e., it is only a function of $q_{x}$ and $q_{y}$. The regular spacing between the BR peaks, which depends on $q_{x}$ and $q_{y}$ only, avoids any possible deformation of these rings due to a sample misalignment. The size of the observed rings does not correspond to the expected from the diameter of the disk. And the intensity outside the first zero in intensity is far higher than the expected from the diffraction of a disk, as deduced from figures \ref{CCD_QyM_CoyFe11} and \ref{Perfiles_FeyCo11}. For instance, the intensity of the BRs at $q_{x}=4q_{0}$ ($q_{0}=\frac{2\pi}{\alpha}$, and $\alpha$ is the lattice parameter of the array) is comparable to the intensity near the region close to the reflected beam, BR $[0,0]$. 

\begin{figure}
\includegraphics[width=8 cm]{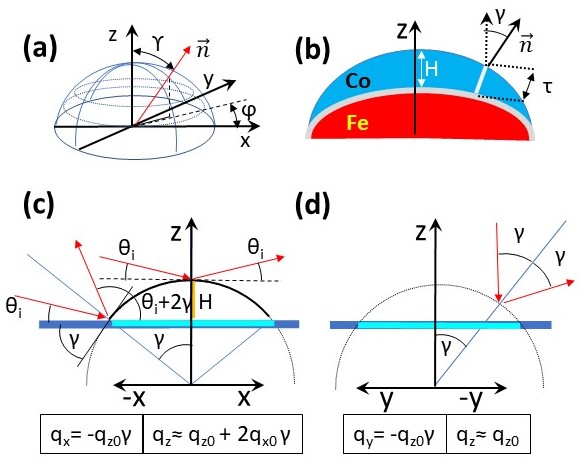}
\caption{Model for the reflection of the x rays from a dome shaped disk. (a) Definition of $\gamma$ and $\varphi$ anges; (b) Transverse cut of the model propossed for the structure of the disks. $H$ is the the thickness of the layer at the center of the disks, and $\tau$ is the thickness at a distance from the center. $\gamma$ is the angle formed by the normal to the surface at that point with respect to the $z$ axis; (c) and (d), dependence of $q_{x}$, $q_{x}$ and $q_{x}$ with the angles $\gamma$ in the longitudinal ($x$ axis) and transverse ($y$ axis) to the beam directions. \label{model}}
\end{figure}

The described intensity distribution in the diffracted patterns is better explained by assuming that the surface of the disks have a curvature due to a radial decrease of their thickness.  This is the only way to obtain the observed particular dispersion of the light along the $q_{x}$ and $q_{y}$ axis and the change in the sign of the magnetic contrast observed in the magnetic dichroism diffraction patterns (figures \ref{CCD_QyM_CoyFe11} and \ref{Perfiles_FeyCo11}).  This change in thickness probably arises by shadowing effects during the deposition of the layers into the antidots. Therefore, the orientation of the surface at each point of the disk at a certain radial distance $\xi$ from the center is characterized by an angle $\gamma$, formed between the normal to the surface in this point and the normal to the sample, and a layer thickness $\tau$. The steepness of the surface should depend on the rate at which $\tau$ decreases, which is unknown. Figure \ref{model} shows the angles and parameters that describe the proposed 3-dimensional shape of the disks.

In this model, the oscillations in intensity are due to the interference between the light scattered at each interface. Therefore, in the case of a single layer, the scattered light with the lowest intensity have the relation $q_{z}\tau=2\pi m+\pi$, where $m$ is an integer number. The thickness $\tau$ is a function of the position in the disk from where the scattered light comes, which depends on $\gamma$. $q_{z}$ is also a function of $\gamma$ due to the way the light is expected to be scattered from the interfaces, which should be done mainly in the same direction than the reflected light. By rotation of the system of reference to align its $z$ axis to the normal of the plane at each point of the disk, it is easy to show that the corresponding moment transfer vector, $\vec{q}$, of the scattered beam has the following variation with the angle $\gamma$ (taking only the linear terms): 

\begin{eqnarray}
q_{x}\approx q_{z0}\gamma\cos\varphi\\
q_{y}\approx q_{z0}\gamma\sin\varphi\\
q_{z}\approx q_{z0}-\frac{q_{x}}{\theta_{i}}
\label{equation02}   
\end{eqnarray}

$q_{z0}=2\left|\vec{k_{i}}\right|\sin{\theta_{i}}$. The angle $\varphi$ is formed by the position of the point in the plane of the disk with respect to the $x$ axis (see figure \ref{model}). These equalities show that $\gamma$ is related to the plane component of the moment transfer vector, $q_{\|}$, by the dependence $\gamma=\frac{q_{\|}}{q_{z0}}$. Note that when $q_{x}=0$, $q_{z}=q_{z0}$ and, therefore, the variation along $q_{y}$ depends only on the variation of the thickness $\tau$. Also, when $q_{x}>0$ ($\cos\varphi>1$), $q_{z}$ decreases, and increases when $q_{x}>0$. This means that the equality $q_{z}\tau=2\pi\left(m+1/2\right)$ is reached at a lower value of $\gamma$ ($q_{x}$) than the corresponding in $[0,q_{y}]$ in the former case, but further away in the $q_{x}<0$. Moreover, due to the higher reflectivity coefficients at grazing angles, it is expected that the scattered intensity decreases as the take off angle of the scattered beam increases, which occurs at $q_{x}<0$ (see figure \ref{geometry}). All of this agrees with what is experimentally observed.

\begin{figure}
\includegraphics[width=8 cm]{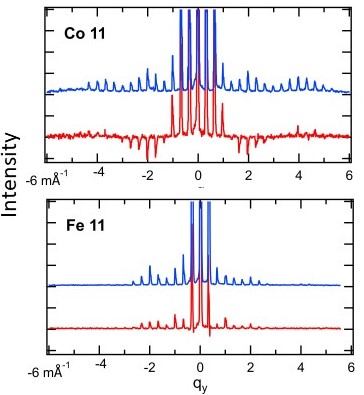}
\caption{Intensity profile along the $q_{y}$ direction at $q_{x}=0$ related to the charge scattering (blue) and the magnetic scattering (red) in cobalt (top) and iron (bottom). \label{CoyFe_11_prfl0qy}}
\end{figure}

Thanks to this result, it is possible to have access to the morphology of the disks in a more quantitative way.  Figure \ref{CoyFe_11_prfl0qy} shows the $[q_{x}=0,q_{y}]$ profiles of the magnetic dichroism and the charge scattering intensities measured in cobalt and iron, in the (11) orientation. This profile allows a better estimation of the change of the thickness of the layers across the disks since the change in $q_{z}$ is practically negligible, all the variations observed are due to the thickness and the steepness of the disk surface. Also, the number of points available are much larger than in the $q_{x}$ direction. In cobalt, the magnetic contrast is perfectly coupled to the charge scattering: the magnetic contrast changes sign every time the charge scattering crosses a point of lowest intensity. This is what expected in the reflection from a single layer, but not from a trilayer system. Actually, the variation in intensity seems to follow a $sin^{2}\left(q_{y}\right)$ function. This suggest that the observed scattered intensity is mainly caused by diffuse scattering at the interfaces of cobalt \cite{2003_XRMS_from_structurally_and_magnetically_rough_interfaces}. This is probably the case since the large incidence angle used ($\theta_{i}=7.7^{\circ}$), which gives rise to a low reflectivity coefficient, the relatively large roughness of the interfaces and the resonant condition. 

The profiles of iron have a similar oscillation period than the found in cobalt, but there is one oscillation less. In this case, its magnetic contrast does not change as in cobalt. In fact, it is more structured in the region of highest intensity, in the central region. There, magnetic contrast changes, but it remains constant in the rest of oscillations. This indicates that the exact understanding of the light scattered from the iron layer is apparently more complicated than in cobalt due to its buried condition. But this complication affects to the magnetic contrast mainly. The oscillation in intensity of the charge scattering is compatible with a single layer model, as in cobalt, which is the most expected behavior at the resonant photon energy. 

Then, the highest order of interference occurs at the point of highest intensity, where the angle $\gamma=0$ and the thickness is the highest. This is $m=3$ for cobalt and $m=2$ for iron due to the lower thickness of the iron layer and the higher wavelength at the iron edge. This agrees with the one less intensity oscillation in iron than in cobalt and the different distribution of the intensity oscillation in the diffraction pattern at both absorption edges. Also, this confirms that the profile along $[0,q_{y}]$ covers scattered light from all the disk, from the highest thickness to the null thickness regions. The dependence of the thickness $\tau$ on $q_{y}$ (which is linear on $\gamma$) is correlated to the rate at which the thickness of the layer changes with the curvature of the surface, which is unknown. For instance, $\tau$ will have a quadratic dependence on $q_{y}$ if the radius of curvature of the surface of the two interfaces is constant. In that case, the zeros in intensity should occur at $q_{y}$ values proportional to the square root of the interference order $m$. However, the experimentally observed relation is close to linear on $m$ (see figure \ref{CoyFe_11_prfl0qy}). This implies that the curvature of the surface of the disks needs to be stronger to have a change in the thickness of the layers. The adjustment of the zeros in the profiles of figure \ref{CoyFe_11_prfl0qy} is done using the equality $q_{z}\tau=2\pi\left(m+1/2\right)$ and taking the relation $\tau= H-\frac{\sigma}{2}\gamma=H-\sigma\frac{q_{y}}{2q_{z0}}$, where $H$ is the highest thickness of the disk, and $\sigma$ a factor that indicates how fast the layer thickness changes with the curvature angle. This shows a slightly faster rate in iron than in cobalt, indicating a larger curvature in cobalt than in iron. This relation is the expected since cobalt is deposited on a curved surface, whereas iron is deposited in a flat surface. As a consequence of this, there is not a perfect one-to-one correspondence in the intensity of the BRs between cobalt and iron. Iron covers a larger area of the disk in a smaller range of $q_{\|}$ than cobalt. This difference is not excessively important. From the previous adjustment of the $[q_{x}=0,q_{y}]$ profile, it is estimated a expansion ratio in cobalt respect to iron of 1.4.

The in-plane component of the moment transfer vector, $\vec{q_{\|}}$, related to the scattered light from the disks will be distorted by a non-uniform curvature of the surface, i.e., by variations in the $\gamma$ angle. This will give rise to a non-uniform distribution of the intensity besides the caused by the thickness. For instance, if the portion of the area of the disk that is flat is large, most of the intensity will be concentrated in a smaller $[q_{x},q_{y}]$ area. Then, although the BR peaks are correlated to the different regions of the disk, i.e., each $[q_{x},q_{y}]$ coordinate is related to a $[x,y]$ position in the disk, this correlation is not completely linear. A precise model of the scattered intensity is required for that, what will determine, therefore, the complete shape of the disks. Note also that the area of the disk covered by the CCD detector is constrained in the $q_{x}$ direction, equivalent to the $x$ direction. This region is delimited by the first interferential zero, meaning that the total decrease in thickness within it is of the order of 60 \AA\ out of 350 \AA.

\section{Hysteresis loops} 
The one-to-one correlation between the in-plane moment transfer vector and the in-plane spatial coordinate of the disks eases the interpretation of the HLs collected at each BR. For instance, this explains why XRMS is specially sensitive to a chiral asymmetry in the disks: when the vortex is formed, the regions of the disks with magnetization parallel to the beam have their normal to their surface mainly pointing transverse to the field. When there is chiral asymmetry in the magnetic vortex circulation, each magnetic orientation points in a direction opposite to the other one giving rise to the resulting asymmetric magnetic contrast in the $q_{y}$ axis. This explanation is different, but not contradictory, to the origin of the sensitivity of XRMS to chiral asymmetry proposed in \cite{chiral_nanotech}, which still holds and it should be observed in perfect flat disks. 

In what follows, it will be assumed that each BR position $[h,k]$ is related to a region around a position $[x,y]$ in the disk. To describe the different regions of the disk, the direction of the incident beam is taken as the reference. This direction is the same than the positive direction of the applied field. Therefore, intensity at $q_{x}>0$ corresponds to the north (N) side of the disk, $q_{x}<0$ to the south (S) side, $q_{y}>0$ to the west (W) side and $q_{y}<0$ to the east (E) side. 

The HLs presented here were normalized to 1. To improve their visualization, their noise was reduced using a binomial smoothing. The smoothing degree was the same for all the loops. This did not modify in essence the loops line-shape since the changes in magnetization should be smooth. However, the smoothing was unable to smearing out all the noise, leaving low frequency oscillations in the magnetization which were obviously more notorious in those loops with poorer signal to noise ratios. Although this did not impede to identify the general trends, it rested accuracy in the value of the onset fields obtained from them, whose highest accuracy is half the field step used to measure the HL, which is 33 Oe. 

Note that the HLs are averaged over hundred of disks. Therefore, the observed result will depend on  the possible number of magnetic configurations that the vortex can adopt. This number obviously decreases when the symmetry of the system decreases, like the one related to the sense of circulation of the magnetization in a vortex.

\begin{figure}
\includegraphics[width=8 cm]{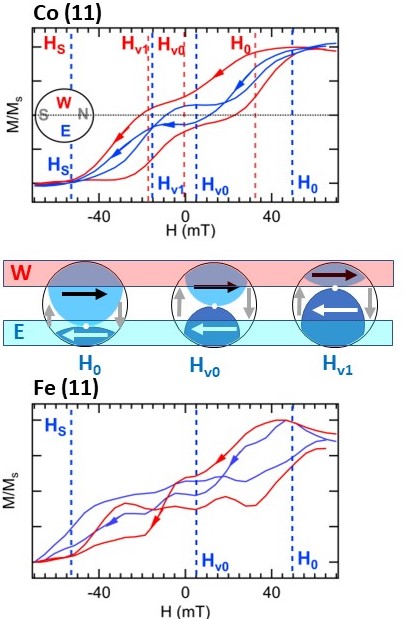}
\caption{On top, HLs collected  at BRs at fixed $q_{x}$ but opposite $q_{y}$ in cobalt at the (11) orientation: in red, $q_{y}>0$; and in blue, $q_{y}<0$. In the middle, a schematics of the related regions of the disks probed at the corresponding BR orientations: $q_{y}>0$ is the related to the W side of the disk, and $q_{y}<0$ to the E side of the disk. The probed regions are the overlaps between the disks and the related stripes, in the corresponding magnetic configuration of the magnetization of the disks at the fields $H_{0}$, $H_{v0}$ and $H_{v1}$. The differences between the HLs are due to a fixed magnetic vortex circulation which is inverted in each branch. In the drawing, N is on the right, S is on the left, W is on the top and E is on the bottom of the disk. The beam direction goes from S to the N direction. On the bottom, the hysteresis loops taken in iron in the (11) orientation and in similar regions than in cobalt. The values of $H_{0}$, $H_{v0}$ and $H_{S}$ signaled in the figure are those of cobalt in the E side HL. \label{vortex_HL}}
\end{figure}

Figure \ref{vortex_HL} displays the HLs observed at the E and W sides of the disk, and located relatively distant from the center, when the applied field was oriented parallel to the (11) orientation of the array. The differences between the two HLs are due to the broken chiral symmetry of the magnetic vortex circulation in this orientation. The HLs shows changes in the magnetic susceptibility at some critical fields which define the onset for the creation, movement and annihilation of the vortex in the disks. The field at which the demagnetization of the disk is initiated is named $H_{0}$. The location of the disk where this happens is of interest since it sets the circulation sense of the vortex. Note that this means that, if the two branches of the loop are symmetrical, the nucleation occurs always in the same side making the sense of circulation in the vortex to invert in each branch. In the presented example, nucleation occurs in the E side, where $H_{0}$ is the highest. Therefore, the circulation is clockwise (CW) in the downward branch and counter clockwise (CCW) in the upward branch. The creation of the vortex causes a fast reduction of the magnetization until a point where the magnetization reaches a value close to zero and the magnetic susceptibility changes again. The field where this occurs is named $H_{v0}$. Once the vortex is formed, the region of the disk with opposite magnetization to the initial one is mainly located in the E side if the circulation is CW. Therefore, its magnetization will be negative at $H_{v0}$, and positive in the W side, being $H_{v0}>0$. The opposite occurs in the upward branch because vortex circulation inverts. This makes the branches of the E side HL to cross each other twice near $\pm H_{v0}$. As the field is increased to magnetized the disk in the opposite direction, the core of the vortex moves transverse to the field. This movement is from the E toward the W in the downward branch. This movement starts at a critical field, named $H_{v1}$. This field has not to be symmetrical to $H_{v0}$. Also, the core movement to the edge might have a different magnetic susceptibility than the changes produced during the creation of the vortex. This makes the  HL to develop lobes near the magnetic saturation regions. The steepness of the magnetic susceptibility in the region between $H_{v0}$ and $H_{v1}$ indicates how much the vortex moves in that range of fields. Therefore, this field region gives direct information of the regions of the disk where core vortex is stable. The HLs of the example shows that the vortex is relatively stable in the region from where the HLs are extracted, what is at the region of the disk far from the center. The killing of the vortex is usually produced near saturation fields. The field at which magnetic saturation is produced is named $H_{s}$. This occurs first in the E side than in the W side. Therefore, the magnetization in the W side is always higher than in the E side in the downward branch, explaining the "fat" shape of its HL, whose branches envelopes those of the E side HL. By contrast, the magnetization measured at any point in the center of the disk from the N to the S sides will be zero if the core of the vortex is in the center of the disk since their main magnetic component is transverse to the measured direction.

\begin{figure*}
\includegraphics[width=18 cm]{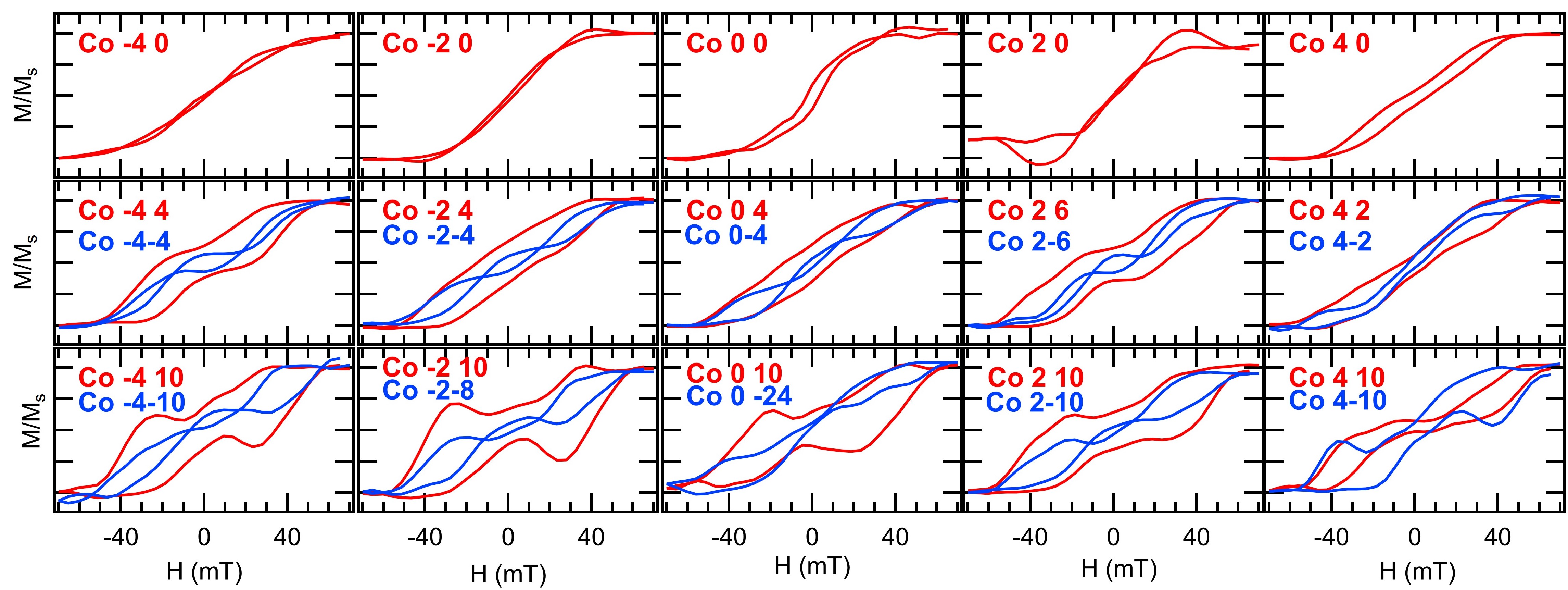}
\caption{Hysteresis loops of the cobalt layer in the (11) orientation, at chosen BRs. The way [h,k] numbers locate the BRs is described in the text. N side is at $q_{x}>0$, E side is at $q_{y}>0$ (red color HLs). \label{Co11_resumen}}
\end{figure*}

\begin{figure*}
\includegraphics[width=18 cm]{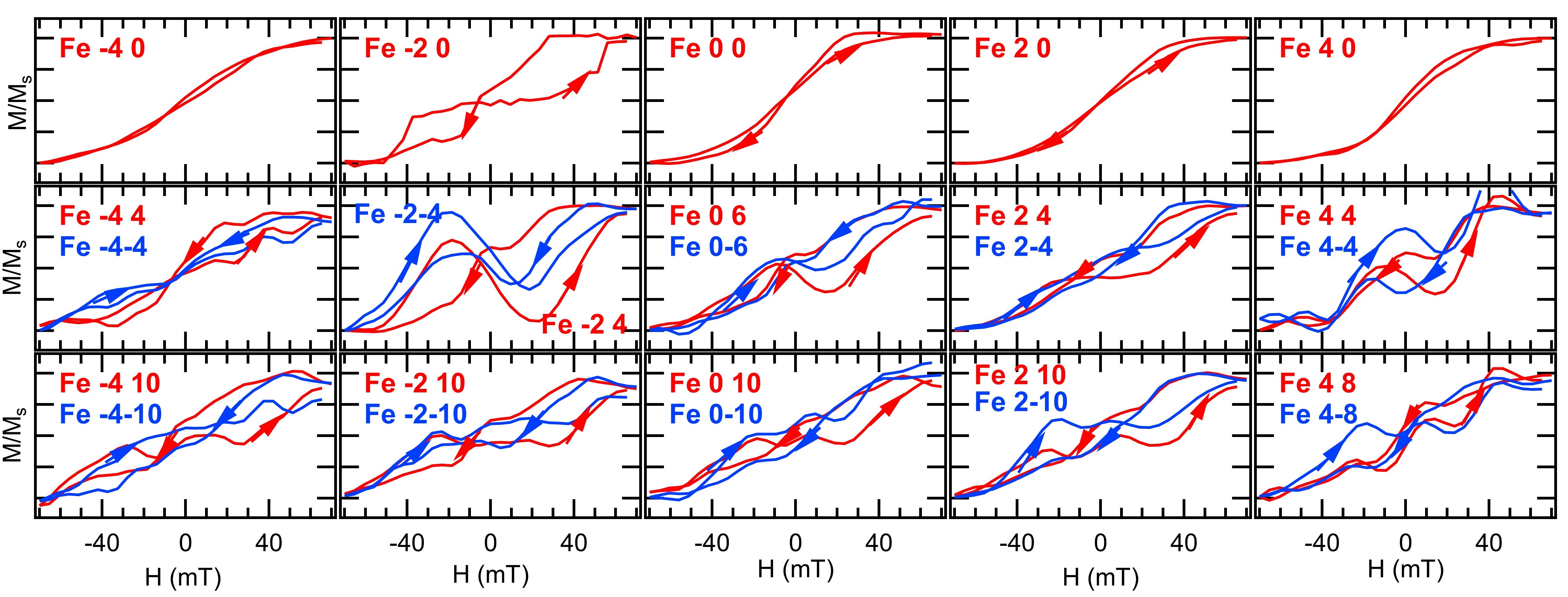}
\caption{Hysteresis loops of the iron layer in the (11) orientation, at chosen BRs. The way [h,k] numbers locate the BRs is described in the text. N side is at $q_{x}>0$, E side is at $q_{y}>0$ (red color HLs).\label{Fe11_resumen}}
\end{figure*}

Figures \ref{Co11_resumen} and \ref{Fe11_resumen} shows the HLs of Fe and Co at different BR positions in the $(11)$ sample orientation, giving a more detailed description of the magnetization at different locations of the disks. The BRs are ordered in the horizontal line from negative $q_{x}$ to positive $q_{x}$ values, which are related to the magnetization at the regions of the disk running from S to N. The first row are the HLs taken  at $q_{y}=0$, i.e., the HLs located at the center of the disk. The other two rows have increasing $q_{y}$  values. They are related to regions of the disk which are increasingly further from the center, either moving towards the E (HLs in blue) or to the W (HLs in red).  The distant $\Delta q$ between BRs along the $q_{x}$ and $q_{y}$ axis is $\sqrt{2}\frac{\pi}{\alpha}$. Some of the $[h,k]$ values displayed in the N side are not exactly the same than at the S side because the corresponding HLs were too distorted to be showed. They are at BRs where the magnetic contrast inverts its sign. The HLs displayed in the third row are taken at $q_{y}$ values that are  further from the center than the allowed $q_{x}$ values ($h\leq 4$).

At the cobalt layer, the HL in BR $[0,0]$ has a coercive field, indicating that the core of the vortex avoids the center of the disk. Note that this HL does not resemble the obtained by VSM (see figure \ref{Kerr_VSM_2Ddisks}). The magnetic behavior in the N and S sides is not symmetric with respect to the center. The $H_{0}$ field is higher in the S side than in the N side. Actually, the HL at the extreme S side does not seem to reach saturation. There, the coercive field is null, but it is significant in the N side. This asymmetry between the N and S sides occurs also in the HLs taken at $q_{y}\neq 0$. The $H_{0}$ field is higher in the S side of the E side (HL in blue), decreasing as $q_{y}$ becomes more negative. The difference in the $H_{0}$ and $H_{s}$ fields between the E (red) and W (blue) side decreases going from the S side to the center. The same behavior occurs from the N side to the center with the important differences that, in this case, the $H_{0}$ is significantly higher in the E side than in the W side. Moreover, the branches of the HL in the NE side never cross each other whereas this clearly happens in the SE side. Therefore, the N side of the disk has a lower probability of holding the core of the vortex than in the S side. The region where the vortex are more stable, i.e., where little changes in the magnetization occurs, is at the edges of the disk (large $\left|{q_{y}}\right|$ values) since it is where the susceptibility between $H_{v0}$ and $H_{v1}$ is flatter and the distance between both fields is increased. It is in the NE and NW sides where the distance between the $H_{v0}$ and $H_{v1}$ fields is the highest, confirming that most of the changes in the vortex in the cobalt layer occurs mainly in the S side.

The HLs of the iron layer shown in figure \ref{Fe11_resumen} are of less quality than those of cobalt due to the lower scattered intensity. All of them have the downward branch different than the upward branch, they are not symmetric. Both branches cross each other once near $H=0$. Figure \ref{vortex_HL} (at the bottom) shows the shape of these HLs more in detail. The start of the downward branch is similar to that of cobalt, which is related to the formation of vortices with the same magnetic circulation sense, CW. The magnetization of the downward branch falls down to lower values at $H\geq$-10 mT. The reduction of magnetization until saturation from there is done with a slow rate. The value of $H_{s}$ is of about 55 mT, similar than the saturation fields found in cobalt. The upward branch starts at the same field $H_{0}$ than the saturation field. This is a much higher $H_{0}$ field than in cobalt. The increase in magnetization is also faster than in the downward branch, stabilizing the vortex at -20 mT. Again, the chirality of this vortex is the same as in cobalt for this orientation of the field, giving raise to a CCW chirality. As in cobalt, the onset field $H_{0}$ is higher in the S side than in the N side. The HL  at [-4,0] has an slope as if magnetic saturation was not complete, something that does not happen at the conjugate BR in [4,0], with lower saturation fields. Therefore, nucleation starts at the same region than in cobalt.

The annihilation of the vortex is obviously different for the two branches. The change in the magnetization at the field where the vortex in the downward branch was killed was more important in the SE side of the disk, indicating that the vortex was preferentially in this region, as in cobalt. Saturation occurs first in the NW side for this branch. For the upward branch, saturation field is higher than in cobalt. Saturation occurs at lower fields in the N side for this loop branch again. In general, saturation is produced first in the NW and at a much higher field in the SE side. The region of the disk where the change in the magnetization in the upward branch between $H_{v0}$ and $H_{v1}$  is small, occurs as well in the edges. But the range of fields seems to be larger towards the S side. From these observations, it seems that the core of vortex in the cobalt and iron layers stays in similar regions of the disks, which could be the reason of the sudden anhinilation of the vortex in the iron layer and the resulting asymmetric HL branches.

\begin{figure*}
\includegraphics[width=18 cm]{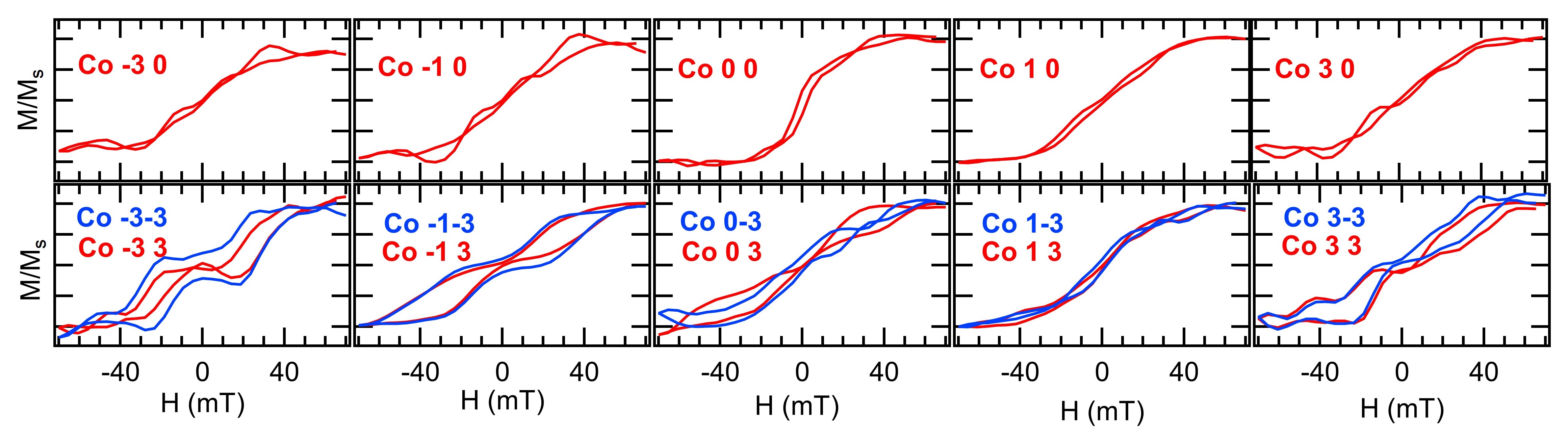}
\caption{Hysteresis loops of the cobalt layer in the (10) orientation, at chosen BRs, [$q_{x}$,$q_{y}$]=[h,k] and in conjugated sites ([h,$\pm$ k]). N side is at $q_{x}>0$, E side is at $q_{y}>0$ (red color HLs).  \label{Co10_resumen}}
\end{figure*}

\begin{figure*}
\includegraphics[width=18 cm]{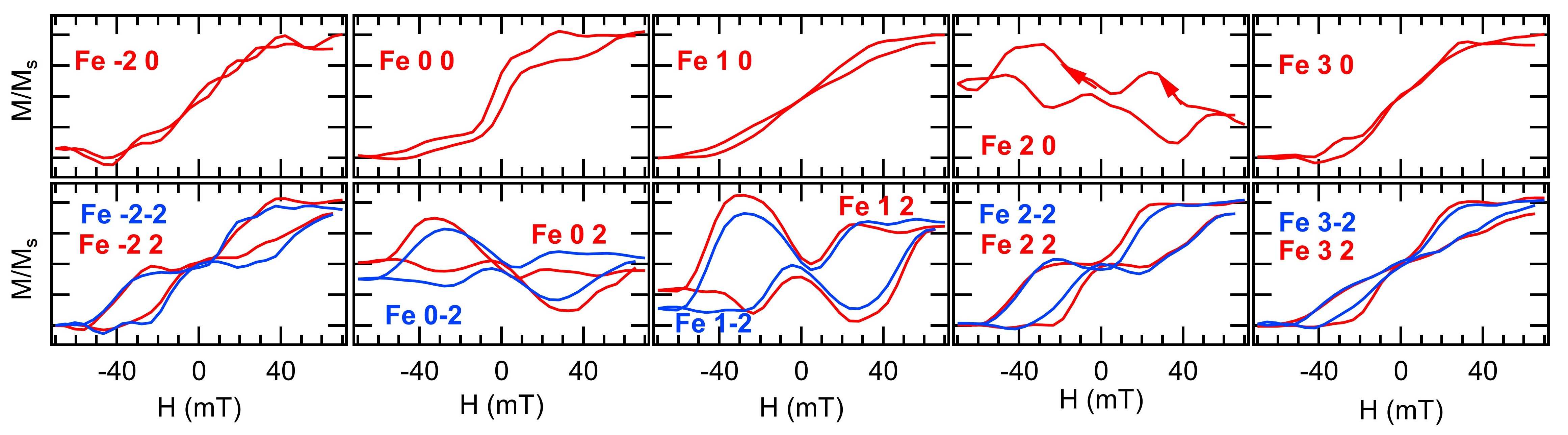}
\caption{Hysteresis loops of the iron layer in the (10) orientation, at chosen BR, [$q_{x}$,$q_{y}$]=[h,k] and in conjugated sites ([h,$\pm$ k]).. N side is at $q_{x}>0$, E side is at $q_{y}>0$ (red color HLs). \label{Fe10_resumen}}
\end{figure*}

Figures \ref{Co10_resumen} and \ref{Fe10_resumen} shows the Fe and Co HLs in the $(10)$ orientation for comparison. In this case, there was not a clear asymmetry in the magnetic vortex circulation. The distance $\Delta q$ between BRs along the $q_{x}$ and $q_{y}$ axis is $\frac{2\pi}{\alpha}$. The HL of the cobalt layer at BR $[0,0]$ is very similar to the obtained in the $(11)$ orientation. The onset field $H_{0}$ is similar in the N and S sides and smaller than in the $(11)$ orientation. Moreover, the highest $H_{0}$ seems to be in the E and W sides, at $[0,\pm 3]$ and $[1,\pm 3]$. There, the HLs have a different shape than at the other edges, indicating that it is in this region where the core of the vortex moves. There is still an asymmetry between the N and the S sides, as in the $(11)$ orientation. At the S side, the HLs with $q_{y}\neq 0$ shows an imbalance in the chiral symmetry of the magnetic circulation. The magnetic circulation is, in this case, CCW in the downward branch and CW in the upward branch. Surprisingly, such a chiral asymmetry is not observed in the N side, remarking the asymmetry in the magnetic behavior between the two sides of the disk. In this side, the downward branch has an onset $H_{0}$ field which is clearly different than in the downward branch. 

In the iron layer, its HL at BR $[0,0]$ has a coercive field which is larger than in cobalt. The onset field $H_{0}$ are similar in the N and S sides. It is also the highest of the registered for this orientation, but smaller than in the $(11)$ orientation. The asymmetry between the N and S sides is not so clear as in cobalt. There is not either a clear imbalance in the chiral symmetry of the vortex. 

\subsection{Discussion}
Having the same chiral asymmetry in both layers is not the expected behavior since the nucleation of the vortex in each layer creates magnetic poles of the same sign. This makes energetically more favorable for each disk to do the nucleation at opposite sides of the border of the disk, inducing an opposite chiral sense in each layer \cite{2008_Magnetic_Vortex}. This is in fact very critical in this case because the region where the disk begins to demagnetize seems to be the same in both layers and for the two orientations of the saturated magnetization in the disks. Therefore, the origin of the asymmetry has to be related to either a higher energy process that should over compensate the polar repulsion between the two layers, or to an effective attractive interaction between layers, possibly induced by the roughness of the interfaces (N\'eel peel orange effect \cite{2000_Neel_orange_peel_coupling}) and/or the observed thickness gradient.

The observed type of chiral asymmetry, which changes of sign depending on the initial magnetization direction, is induced in submicron magnets by making their shape non-centrosymetric: triangles, truncated disks, asymmetric rings, "pac-man" shaped disks or asymmetrical magnetic moment distribution\cite{2001_Magnetic_switching,2010_Control_of_Double_Vortex, 2011_Chirality_control, 2010_Control_of_the_chirality,2014_Controlling_vortex_chirality, 2011_Control_of_vortex_chirality, 2009_Vortex_chirality_control}. Chiral asymmetry occurs in these systems when the applied field is at an angle with respect to the mirror symmetry axis of the system.

\begin{figure}
\includegraphics[width=8 cm]{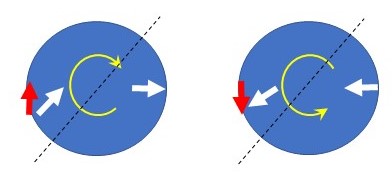}
\caption{Model proposed to explain the chiral asymmetry in the array of disks. \label{model_chirality}}
\end{figure}

To explain the origin of the chiral asymmetry in the studied system, we proposed the following model based on the existence of a non-centrosymmetry in the magnets of the array. In the $(10)$ field orientation, the cobalt layer contains an asymmetry.Such an asymmetry might be related to the anisotropy energy which could be asymmetrically distributed across the area of the disk. This could happen if its shape is not fully symmetric. In this case, the S side has possibly a higher anisotropy than  the N side, what is required to preserve mirror symmetry between the W and E sides, since chiral asymmetry is not fully manifested in this orientation \cite{2014_Controlling_vortex_chirality}. In the $(11)$ orientation, the oblique angle direction of the applied field breaks such a mirror symmetry causing the observed chiral asymmetry. Then, when the direction of the field is positive, the moments in the N side have a weaker anisotropy and align their moments with the applied field at lower fields than in the S side. This creates an in-plane component of the magnetization perpendicular to the direction of the field at that particular region, setting the sense of rotation of the magnetization in the vortex. When the direction of the field is the opposite, the orientation of this in-plane perpendicular component changes its direction as well, changing the sense of the magnetic circulation of the vortex to the opposite one. The model is depicted in figure \ref{model_chirality}. This process seems to be very solid in the cobalt layer since the branches of its HLs are symmetric at any point in the disk. It is also the layer that register an asymmetry between N a S sides in the (10) orientation as well.

\begin{figure}
\includegraphics[width=8 cm]{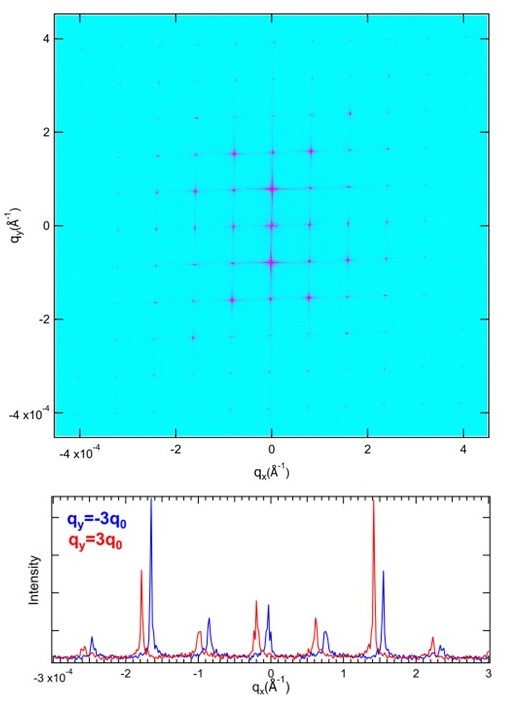}
\caption{Fourier transform of a SEM image of the array including more than $10^{3}$ disks. On the bottom, the intensity of $q_{x}$ scans across the previous image in two opossite values of $q_{y}$ to evidence the deviation from perfect  mirror symmetry with respect to the (10) and (01) axes. Scans have their $q_{x}$ zero shifted for better comparison. $q_{0}=2\pi/\alpha$ where $\alpha$ is the lattice constant of the array \label{FFT_SEM}}
\end{figure}

Figure \ref{FFT_SEM} shows the Fast Fourier Transform (FFT) of the image obtained by SEM of the sample which included more than $10^{3}$ disks. A visual inspection shows that the intensity of the peaks deviates from the expected symmetric square shape resulting from the FFT of a diamond shape. A detailed analysis of the intensity of the BR peaks in this figure shows that the ideally diamond-shape disks are actually rhomboids. Their shape asymmetry arises because the distance between opposite sides of the diamond-like shape of the disks are not exactly the same. This makes the shape of the disks non-centrosymmetric, i.e., there is not perfect mirror symmetry across any of the symmetry axis oriented along the (10), (01), (11) or (-1-1) directions. This deviation from centrosymmetry in the shape of the magnets is much smaller than the non-centrosymmetry induced in magnets by purpose to fix their vortex chirality. This might indicate that a precise control of the ILL technique can be used to modulate the chiral properties of magnets arrays by inducing asymmetries in their shape.

Since the HL measurement was done in a single cycle, it is not possible to assert that the different magnetization paths taken in each of the branches of the HL in Fe in the (11) orientation, and the observed in some HLs in the (10) orientation, was a systematic and repetitive process. The HL measured by VSM, which covers a much larger area than the measured by XRMS, did not show any asymmetry between the branches. Therefore, these asymmetries are possibly due to a certain stochasticity in the magnetic inversion process of the disks, which have to be collective by the nature of the measurement. 

\section{Conclusions}
In conclusion, we observed a one-to-one correlation between the in-plane moment vector transfer of the scattered light from an array of submicron magnets and the spatial location in the submicron  magnet from where the light comes, converting the magnetic contrast of the diffraction pattern of the array in an image of the magnetization of the disk, averaged over the illuminated disks. The conditions for this to happen are related to the morphology of the magnets and the configuration of the experiment. The surface of the magnets must have a certain curvature. In the present experiment, this is obtained using magnetron sputtering deposition at normal incidence in patterned holes, which is one of the most used methods to produce the kind of studied submicron magnet arrays. The grazing incidence angles employed were relatively large to allow the collection of scattered light at a wider range of in-plane moment transfer. At such large angles, the diffuse scattering due to the roughness and imperfections at the layer interfaces should be important what, joined to the resonant energies employed, allowed a better isolation of the targeted layers even if they were buried under 20 nm thick layers. This kind of magnetic microscope effect explains why XRMS is specially sensitive to the chiral asymmetry of the magnetization circulation of the vortex in this kind of samples. This was used to study the chiral asymmetry of the disks. Thanks to this effect, it was possible to identify the presence of a non-centro-symmetry in the magnetization of the sample that explained the apparition of the chiral asymmetry at the oblique angle orientation of the field with respect to the EA axis of the array. The physical origin of such a magnetic non-centro-symmetry was attributed to deviations from centro-symmetry in the shape of the magnets, giving an indication of the sensitivity of the studied system to such deviations. The presented experiment shows that XRMS can give a collective vision of the stability and symmetry breaking process in this kind of system which is complementary to the obtained by other microscopic techniques. There is plenty of room to increase the quality of the data and to increase the information extracted from the scattered light in the configuration used in this experiment, specially the related to the morphology of the magnets. 

\begin{acknowledgments}
This project has been supported by Spanish MINECO under grant PID2019-104604RB/AEI/10.13039/501100011033, and by Asturias FICYT under grant AYUD/2021/51185 with the support of FEDER funds. R. M. acknowledge Basque Country grant No. IT1491-22.
\end{acknowledgments}


\begin{thebibliography}{41}%
\makeatletter
\providecommand \@ifxundefined [1]{%
 \@ifx{#1\undefined}
}%
\providecommand \@ifnum [1]{%
 \ifnum #1\expandafter \@firstoftwo
 \else \expandafter \@secondoftwo
 \fi
}%
\providecommand \@ifx [1]{%
 \ifx #1\expandafter \@firstoftwo
 \else \expandafter \@secondoftwo
 \fi
}%
\providecommand \natexlab [1]{#1}%
\providecommand \enquote  [1]{``#1''}%
\providecommand \bibnamefont  [1]{#1}%
\providecommand \bibfnamefont [1]{#1}%
\providecommand \citenamefont [1]{#1}%
\providecommand \href@noop [0]{\@secondoftwo}%
\providecommand \href [0]{\begingroup \@sanitize@url \@href}%
\providecommand \@href[1]{\@@startlink{#1}\@@href}%
\providecommand \@@href[1]{\endgroup#1\@@endlink}%
\providecommand \@sanitize@url [0]{\catcode `\\12\catcode `\$12\catcode
  `\&12\catcode `\#12\catcode `\^12\catcode `\_12\catcode `\%12\relax}%
\providecommand \@@startlink[1]{}%
\providecommand \@@endlink[0]{}%
\providecommand \url  [0]{\begingroup\@sanitize@url \@url }%
\providecommand \@url [1]{\endgroup\@href {#1}{\urlprefix }}%
\providecommand \urlprefix  [0]{URL }%
\providecommand \Eprint [0]{\href }%
\providecommand \doibase [0]{https://doi.org/}%
\providecommand \selectlanguage [0]{\@gobble}%
\providecommand \bibinfo  [0]{\@secondoftwo}%
\providecommand \bibfield  [0]{\@secondoftwo}%
\providecommand \translation [1]{[#1]}%
\providecommand \BibitemOpen [0]{}%
\providecommand \bibitemStop [0]{}%
\providecommand \bibitemNoStop [0]{.\EOS\space}%
\providecommand \EOS [0]{\spacefactor3000\relax}%
\providecommand \BibitemShut  [1]{\csname bibitem#1\endcsname}%
\let\auto@bib@innerbib\@empty
\bibitem [{\citenamefont {Guslienko}\ \emph {et~al.}(2001)\citenamefont
  {Guslienko}, \citenamefont {Novosad}, \citenamefont {Otani}, \citenamefont
  {Shima},\ and\ \citenamefont {Fukamichi}}]{2001_Field_evolution}%
  \BibitemOpen
  \bibfield  {author} {\bibinfo {author} {\bibfnamefont {K.~Y.}\ \bibnamefont
  {Guslienko}}, \bibinfo {author} {\bibfnamefont {V.}~\bibnamefont {Novosad}},
  \bibinfo {author} {\bibfnamefont {Y.}~\bibnamefont {Otani}}, \bibinfo
  {author} {\bibfnamefont {H.}~\bibnamefont {Shima}},\ and\ \bibinfo {author}
  {\bibfnamefont {K.}~\bibnamefont {Fukamichi}},\ }\href
  {https://doi.org/10.1063/1.1377850} {\bibfield  {journal} {\bibinfo
  {journal} {Applied Physics Letters}\ }\textbf {\bibinfo {volume} {78}},\
  \bibinfo {pages} {3848} (\bibinfo {year} {2001})}\BibitemShut {NoStop}%
\bibitem [{\citenamefont {Schneider}\ \emph {et~al.}(2001)\citenamefont
  {Schneider}, \citenamefont {Hoffmann},\ and\ \citenamefont
  {Zweck}}]{2001_Magnetic_switching}%
  \BibitemOpen
  \bibfield  {author} {\bibinfo {author} {\bibfnamefont {M.}~\bibnamefont
  {Schneider}}, \bibinfo {author} {\bibfnamefont {H.}~\bibnamefont
  {Hoffmann}},\ and\ \bibinfo {author} {\bibfnamefont {J.}~\bibnamefont
  {Zweck}},\ }\href {https://doi.org/10.1063/1.1410873} {\bibfield  {journal}
  {\bibinfo  {journal} {Applied Physics Letters}\ }\textbf {\bibinfo {volume}
  {79}},\ \bibinfo {pages} {3113} (\bibinfo {year} {2001})}\BibitemShut
  {NoStop}%
\bibitem [{\citenamefont {Grimsditch}\ \emph {et~al.}(2002)\citenamefont
  {Grimsditch}, \citenamefont {Vavassori}, \citenamefont {Novosad},
  \citenamefont {Metlushko}, \citenamefont {Shima}, \citenamefont {Otani},\
  and\ \citenamefont {Fukamichi}}]{2002_Vortex_chirality}%
  \BibitemOpen
  \bibfield  {author} {\bibinfo {author} {\bibfnamefont {M.}~\bibnamefont
  {Grimsditch}}, \bibinfo {author} {\bibfnamefont {P.}~\bibnamefont
  {Vavassori}}, \bibinfo {author} {\bibfnamefont {V.}~\bibnamefont {Novosad}},
  \bibinfo {author} {\bibfnamefont {V.}~\bibnamefont {Metlushko}}, \bibinfo
  {author} {\bibfnamefont {H.}~\bibnamefont {Shima}}, \bibinfo {author}
  {\bibfnamefont {Y.}~\bibnamefont {Otani}},\ and\ \bibinfo {author}
  {\bibfnamefont {K.}~\bibnamefont {Fukamichi}},\ }\href
  {https://doi.org/10.1103/PhysRevB.65.172419} {\bibfield  {journal} {\bibinfo
  {journal} {Phys. Rev. B}\ }\textbf {\bibinfo {volume} {65}},\ \bibinfo
  {pages} {172419} (\bibinfo {year} {2002})}\BibitemShut {NoStop}%
\bibitem [{\citenamefont {Buchanan}\ \emph {et~al.}(2005)\citenamefont
  {Buchanan}, \citenamefont {Guslienko}, \citenamefont {Doran}, \citenamefont
  {Scholl}, \citenamefont {Bader},\ and\ \citenamefont
  {Novosad}}]{2005_Magnetic_remanent}%
  \BibitemOpen
  \bibfield  {author} {\bibinfo {author} {\bibfnamefont {K.~S.}\ \bibnamefont
  {Buchanan}}, \bibinfo {author} {\bibfnamefont {K.~Y.}\ \bibnamefont
  {Guslienko}}, \bibinfo {author} {\bibfnamefont {A.}~\bibnamefont {Doran}},
  \bibinfo {author} {\bibfnamefont {A.}~\bibnamefont {Scholl}}, \bibinfo
  {author} {\bibfnamefont {S.~D.}\ \bibnamefont {Bader}},\ and\ \bibinfo
  {author} {\bibfnamefont {V.}~\bibnamefont {Novosad}},\ }\href
  {https://doi.org/10.1103/PhysRevB.72.134415} {\bibfield  {journal} {\bibinfo
  {journal} {Phys. Rev. B}\ }\textbf {\bibinfo {volume} {72}},\ \bibinfo
  {pages} {134415} (\bibinfo {year} {2005})}\BibitemShut {NoStop}%
\bibitem [{\citenamefont {Guslienko}\ and\ \citenamefont
  {Division}(2008)}]{2008_Magnetic_Vortex}%
  \BibitemOpen
  \bibfield  {author} {\bibinfo {author} {\bibfnamefont {K.~Y.}\ \bibnamefont
  {Guslienko}}\ and\ \bibinfo {author} {\bibfnamefont {M.~S.}\ \bibnamefont
  {Division}},\ }\href {https://www.osti.gov/biblio/935275} {\bibfield
  {journal} {\bibinfo  {journal} {J. Nanosci. Nanotechnol.}\ }\textbf {\bibinfo
  {volume} {8}},\ \bibinfo {pages} {2745} (\bibinfo {year} {2008})}\BibitemShut
  {NoStop}%
\bibitem [{\citenamefont {Zhu}\ \emph {et~al.}(2000)\citenamefont {Zhu},
  \citenamefont {Zheng},\ and\ \citenamefont {Prinz}}]{2000_Ultrahigh_density}%
  \BibitemOpen
  \bibfield  {author} {\bibinfo {author} {\bibfnamefont {J.-G.}\ \bibnamefont
  {Zhu}}, \bibinfo {author} {\bibfnamefont {Y.}~\bibnamefont {Zheng}},\ and\
  \bibinfo {author} {\bibfnamefont {G.~A.}\ \bibnamefont {Prinz}},\ }\href
  {https://doi.org/10.1063/1.372805} {\bibfield  {journal} {\bibinfo  {journal}
  {Journal of Applied Physics}\ }\textbf {\bibinfo {volume} {87}},\ \bibinfo
  {pages} {6668} (\bibinfo {year} {2000})}\BibitemShut {NoStop}%
\bibitem [{\citenamefont {Pigeau}\ \emph {et~al.}(2010)\citenamefont {Pigeau},
  \citenamefont {de~Loubens}, \citenamefont {Klein}, \citenamefont {Riegler},
  \citenamefont {Lochner}, \citenamefont {Schmidt}, \citenamefont {Molenkamp},
  \citenamefont {Tiberkevich},\ and\ \citenamefont
  {Slavin}}]{2010_A_frequency_controlled}%
  \BibitemOpen
  \bibfield  {author} {\bibinfo {author} {\bibfnamefont {B.}~\bibnamefont
  {Pigeau}}, \bibinfo {author} {\bibfnamefont {G.}~\bibnamefont {de~Loubens}},
  \bibinfo {author} {\bibfnamefont {O.}~\bibnamefont {Klein}}, \bibinfo
  {author} {\bibfnamefont {A.}~\bibnamefont {Riegler}}, \bibinfo {author}
  {\bibfnamefont {F.}~\bibnamefont {Lochner}}, \bibinfo {author} {\bibfnamefont
  {G.}~\bibnamefont {Schmidt}}, \bibinfo {author} {\bibfnamefont {L.~W.}\
  \bibnamefont {Molenkamp}}, \bibinfo {author} {\bibfnamefont {V.~S.}\
  \bibnamefont {Tiberkevich}},\ and\ \bibinfo {author} {\bibfnamefont {A.~N.}\
  \bibnamefont {Slavin}},\ }\href {https://doi.org/10.1063/1.3373833}
  {\bibfield  {journal} {\bibinfo  {journal} {Applied Physics Letters}\
  }\textbf {\bibinfo {volume} {96}},\ \bibinfo {pages} {132506} (\bibinfo
  {year} {2010})}\BibitemShut {NoStop}%
\bibitem [{\citenamefont {Guslienko}\ \emph {et~al.}(2005)\citenamefont
  {Guslienko}, \citenamefont {Buchanan}, \citenamefont {Bader},\ and\
  \citenamefont {Novosad}}]{2005_Dynamics_of_coupled}%
  \BibitemOpen
  \bibfield  {author} {\bibinfo {author} {\bibfnamefont {K.~Y.}\ \bibnamefont
  {Guslienko}}, \bibinfo {author} {\bibfnamefont {K.~S.}\ \bibnamefont
  {Buchanan}}, \bibinfo {author} {\bibfnamefont {S.~D.}\ \bibnamefont
  {Bader}},\ and\ \bibinfo {author} {\bibfnamefont {V.}~\bibnamefont
  {Novosad}},\ }\href {https://doi.org/10.1063/1.1929078} {\bibfield  {journal}
  {\bibinfo  {journal} {Applied Physics Letters}\ }\textbf {\bibinfo {volume}
  {86}},\ \bibinfo {pages} {223112} (\bibinfo {year} {2005})}\BibitemShut
  {NoStop}%
\bibitem [{\citenamefont {Wintz}\ \emph {et~al.}(2016)\citenamefont {Wintz},
  \citenamefont {Tiberkevich}, \citenamefont {Weigand}, \citenamefont {Raabe},
  \citenamefont {Lindner}, \citenamefont {Erbe}, \citenamefont {Slavin},\ and\
  \citenamefont {Fassbender}}]{2016_Magnetic_vortex_cores}%
  \BibitemOpen
  \bibfield  {author} {\bibinfo {author} {\bibfnamefont {S.}~\bibnamefont
  {Wintz}}, \bibinfo {author} {\bibfnamefont {V.}~\bibnamefont {Tiberkevich}},
  \bibinfo {author} {\bibfnamefont {M.}~\bibnamefont {Weigand}}, \bibinfo
  {author} {\bibfnamefont {J.}~\bibnamefont {Raabe}}, \bibinfo {author}
  {\bibfnamefont {J.}~\bibnamefont {Lindner}}, \bibinfo {author} {\bibfnamefont
  {A.}~\bibnamefont {Erbe}}, \bibinfo {author} {\bibfnamefont {A.}~\bibnamefont
  {Slavin}},\ and\ \bibinfo {author} {\bibfnamefont {J.}~\bibnamefont
  {Fassbender}},\ }\href {https://doi.org/10.1038/nnano.2016.117} {\bibfield
  {journal} {\bibinfo  {journal} {Nature Nanotechnology}\ }\textbf {\bibinfo
  {volume} {11}},\ \bibinfo {pages} {948} (\bibinfo {year} {2016})}\BibitemShut
  {NoStop}%
\bibitem [{\citenamefont {Suess}\ \emph {et~al.}(2018)\citenamefont {Suess},
  \citenamefont {Bachleitner-Hofmann}, \citenamefont {Satz}, \citenamefont
  {Weitensfelder}, \citenamefont {Vogler}, \citenamefont {Bruckner},
  \citenamefont {Abert}, \citenamefont {Pr{\"u}gl}, \citenamefont {Zimmer},
  \citenamefont {Huber}, \citenamefont {Luber}, \citenamefont {Raberg},
  \citenamefont {Schrefl},\ and\ \citenamefont
  {Br{\"u}ckl}}]{2018_Topologically_protected}%
  \BibitemOpen
  \bibfield  {author} {\bibinfo {author} {\bibfnamefont {D.}~\bibnamefont
  {Suess}}, \bibinfo {author} {\bibfnamefont {A.}~\bibnamefont
  {Bachleitner-Hofmann}}, \bibinfo {author} {\bibfnamefont {A.}~\bibnamefont
  {Satz}}, \bibinfo {author} {\bibfnamefont {H.}~\bibnamefont {Weitensfelder}},
  \bibinfo {author} {\bibfnamefont {C.}~\bibnamefont {Vogler}}, \bibinfo
  {author} {\bibfnamefont {F.}~\bibnamefont {Bruckner}}, \bibinfo {author}
  {\bibfnamefont {C.}~\bibnamefont {Abert}}, \bibinfo {author} {\bibfnamefont
  {K.}~\bibnamefont {Pr{\"u}gl}}, \bibinfo {author} {\bibfnamefont
  {J.}~\bibnamefont {Zimmer}}, \bibinfo {author} {\bibfnamefont
  {C.}~\bibnamefont {Huber}}, \bibinfo {author} {\bibfnamefont
  {S.}~\bibnamefont {Luber}}, \bibinfo {author} {\bibfnamefont
  {W.}~\bibnamefont {Raberg}}, \bibinfo {author} {\bibfnamefont
  {T.}~\bibnamefont {Schrefl}},\ and\ \bibinfo {author} {\bibfnamefont
  {H.}~\bibnamefont {Br{\"u}ckl}},\ }\href
  {https://doi.org/10.1038/s41928-018-0084-2} {\bibfield  {journal} {\bibinfo
  {journal} {Nature Electronics}\ }\textbf {\bibinfo {volume} {1}},\ \bibinfo
  {pages} {362} (\bibinfo {year} {2018})}\BibitemShut {NoStop}%
\bibitem [{\citenamefont {Endo}\ \emph {et~al.}(2022)\citenamefont {Endo},
  \citenamefont {Al-Mahdawi}, \citenamefont {Oogane},\ and\ \citenamefont
  {Ando}}]{2022_Control_of_sensitivity}%
  \BibitemOpen
  \bibfield  {author} {\bibinfo {author} {\bibfnamefont {M.}~\bibnamefont
  {Endo}}, \bibinfo {author} {\bibfnamefont {M.}~\bibnamefont {Al-Mahdawi}},
  \bibinfo {author} {\bibfnamefont {M.}~\bibnamefont {Oogane}},\ and\ \bibinfo
  {author} {\bibfnamefont {Y.}~\bibnamefont {Ando}},\ }\href
  {https://doi.org/10.1088/1361-6463/ac5080} {\bibfield  {journal} {\bibinfo
  {journal} {Journal of Physics D: Applied Physics}\ }\textbf {\bibinfo
  {volume} {55}},\ \bibinfo {pages} {195001} (\bibinfo {year}
  {2022})}\BibitemShut {NoStop}%
\bibitem [{\citenamefont {Wurft}\ \emph {et~al.}(2019)\citenamefont {Wurft},
  \citenamefont {Raberg}, \citenamefont {Prügl}, \citenamefont {Satz},
  \citenamefont {Reiss},\ and\ \citenamefont {Brückl}}]{2019_Evolution}%
  \BibitemOpen
  \bibfield  {author} {\bibinfo {author} {\bibfnamefont {T.}~\bibnamefont
  {Wurft}}, \bibinfo {author} {\bibfnamefont {W.}~\bibnamefont {Raberg}},
  \bibinfo {author} {\bibfnamefont {K.}~\bibnamefont {Prügl}}, \bibinfo
  {author} {\bibfnamefont {A.}~\bibnamefont {Satz}}, \bibinfo {author}
  {\bibfnamefont {G.}~\bibnamefont {Reiss}},\ and\ \bibinfo {author}
  {\bibfnamefont {H.}~\bibnamefont {Brückl}},\ }\href
  {https://doi.org/10.1063/1.5116299} {\bibfield  {journal} {\bibinfo
  {journal} {Applied Physics Letters}\ }\textbf {\bibinfo {volume} {115}},\
  \bibinfo {pages} {132407} (\bibinfo {year} {2019})}\BibitemShut {NoStop}%
\bibitem [{\citenamefont {Kim}\ \emph {et~al.}(2010)\citenamefont {Kim},
  \citenamefont {Rozhkova}, \citenamefont {Ulasov}, \citenamefont {Bader},
  \citenamefont {Rajh}, \citenamefont {Lesniak},\ and\ \citenamefont
  {Novosad}}]{2009_Biofunctionalized_magnetic_vortex}%
  \BibitemOpen
  \bibfield  {author} {\bibinfo {author} {\bibfnamefont {D.-H.}\ \bibnamefont
  {Kim}}, \bibinfo {author} {\bibfnamefont {E.~A.}\ \bibnamefont {Rozhkova}},
  \bibinfo {author} {\bibfnamefont {I.~V.}\ \bibnamefont {Ulasov}}, \bibinfo
  {author} {\bibfnamefont {S.~D.}\ \bibnamefont {Bader}}, \bibinfo {author}
  {\bibfnamefont {T.}~\bibnamefont {Rajh}}, \bibinfo {author} {\bibfnamefont
  {M.~S.}\ \bibnamefont {Lesniak}},\ and\ \bibinfo {author} {\bibfnamefont
  {V.}~\bibnamefont {Novosad}},\ }\href {https://doi.org/10.1038/nmat2591}
  {\bibfield  {journal} {\bibinfo  {journal} {Nature Materials}\ }\textbf
  {\bibinfo {volume} {9}},\ \bibinfo {pages} {165} (\bibinfo {year}
  {2010})}\BibitemShut {NoStop}%
\bibitem [{\citenamefont {Peixoto}\ \emph {et~al.}(2020)\citenamefont
  {Peixoto}, \citenamefont {Magalhães}, \citenamefont {Navas}, \citenamefont
  {Moraes}, \citenamefont {Redondo}, \citenamefont {Morales}, \citenamefont
  {Araújo},\ and\ \citenamefont {Sousa}}]{2020_Magnetic_nanostructures}%
  \BibitemOpen
  \bibfield  {author} {\bibinfo {author} {\bibfnamefont {L.}~\bibnamefont
  {Peixoto}}, \bibinfo {author} {\bibfnamefont {R.}~\bibnamefont {Magalhães}},
  \bibinfo {author} {\bibfnamefont {D.}~\bibnamefont {Navas}}, \bibinfo
  {author} {\bibfnamefont {S.}~\bibnamefont {Moraes}}, \bibinfo {author}
  {\bibfnamefont {C.}~\bibnamefont {Redondo}}, \bibinfo {author} {\bibfnamefont
  {R.}~\bibnamefont {Morales}}, \bibinfo {author} {\bibfnamefont {J.~P.}\
  \bibnamefont {Araújo}},\ and\ \bibinfo {author} {\bibfnamefont {C.~T.}\
  \bibnamefont {Sousa}},\ }\href {https://doi.org/10.1063/1.5121702} {\bibfield
   {journal} {\bibinfo  {journal} {Applied Physics Reviews}\ }\textbf {\bibinfo
  {volume} {7}},\ \bibinfo {pages} {011310} (\bibinfo {year}
  {2020})}\BibitemShut {NoStop}%
\bibitem [{\citenamefont {Huang}\ \emph {et~al.}(2010)\citenamefont {Huang},
  \citenamefont {Schofield},\ and\ \citenamefont
  {Zhu}}]{2010_Control_of_Double_Vortex}%
  \BibitemOpen
  \bibfield  {author} {\bibinfo {author} {\bibfnamefont {L.}~\bibnamefont
  {Huang}}, \bibinfo {author} {\bibfnamefont {M.~A.}\ \bibnamefont
  {Schofield}},\ and\ \bibinfo {author} {\bibfnamefont {Y.}~\bibnamefont
  {Zhu}},\ }\href {https://doi.org/https://doi.org/10.1002/adma.200902488}
  {\bibfield  {journal} {\bibinfo  {journal} {Advanced Materials}\ }\textbf
  {\bibinfo {volume} {22}},\ \bibinfo {pages} {492} (\bibinfo {year}
  {2010})}\BibitemShut {NoStop}%
\bibitem [{\citenamefont {Vavassori}\ \emph {et~al.}(2008)\citenamefont
  {Vavassori}, \citenamefont {Bonanni}, \citenamefont {Busato}, \citenamefont
  {Bisero}, \citenamefont {Gubbiotti}, \citenamefont {Adeyeye}, \citenamefont
  {Goolaup}, \citenamefont {Singh}, \citenamefont {Spezzani},\ and\
  \citenamefont {Sacchi}}]{2008_Magnetostatic_and_exchange}%
  \BibitemOpen
  \bibfield  {author} {\bibinfo {author} {\bibfnamefont {P.}~\bibnamefont
  {Vavassori}}, \bibinfo {author} {\bibfnamefont {V.}~\bibnamefont {Bonanni}},
  \bibinfo {author} {\bibfnamefont {A.}~\bibnamefont {Busato}}, \bibinfo
  {author} {\bibfnamefont {D.}~\bibnamefont {Bisero}}, \bibinfo {author}
  {\bibfnamefont {G.}~\bibnamefont {Gubbiotti}}, \bibinfo {author}
  {\bibfnamefont {A.~O.}\ \bibnamefont {Adeyeye}}, \bibinfo {author}
  {\bibfnamefont {S.}~\bibnamefont {Goolaup}}, \bibinfo {author} {\bibfnamefont
  {N.}~\bibnamefont {Singh}}, \bibinfo {author} {\bibfnamefont
  {C.}~\bibnamefont {Spezzani}},\ and\ \bibinfo {author} {\bibfnamefont
  {M.}~\bibnamefont {Sacchi}},\ }\href
  {https://doi.org/10.1088/0022-3727/41/13/134014} {\bibfield  {journal}
  {\bibinfo  {journal} {Journal of Physics D: Applied Physics}\ }\textbf
  {\bibinfo {volume} {41}},\ \bibinfo {pages} {134014} (\bibinfo {year}
  {2008})}\BibitemShut {NoStop}%
\bibitem [{\citenamefont {Parente}\ \emph {et~al.}(2022)\citenamefont
  {Parente}, \citenamefont {Navarro}, \citenamefont {Vargas}, \citenamefont
  {Lapa}, \citenamefont {Basaran}, \citenamefont {González}, \citenamefont
  {Redondo}, \citenamefont {Morales}, \citenamefont {Munoz~Noval},
  \citenamefont {Schuller},\ and\ \citenamefont
  {Vicent}}]{2022_Unusual_Magnetic_Hysteresis_and_Transition_between_Vortex_and_Double_Pole_States}%
  \BibitemOpen
  \bibfield  {author} {\bibinfo {author} {\bibfnamefont {A.}~\bibnamefont
  {Parente}}, \bibinfo {author} {\bibfnamefont {H.}~\bibnamefont {Navarro}},
  \bibinfo {author} {\bibfnamefont {N.~M.}\ \bibnamefont {Vargas}}, \bibinfo
  {author} {\bibfnamefont {P.}~\bibnamefont {Lapa}}, \bibinfo {author}
  {\bibfnamefont {A.~C.}\ \bibnamefont {Basaran}}, \bibinfo {author}
  {\bibfnamefont {E.~M.}\ \bibnamefont {González}}, \bibinfo {author}
  {\bibfnamefont {C.}~\bibnamefont {Redondo}}, \bibinfo {author} {\bibfnamefont
  {R.}~\bibnamefont {Morales}}, \bibinfo {author} {\bibfnamefont
  {A.}~\bibnamefont {Munoz~Noval}}, \bibinfo {author} {\bibfnamefont {I.~K.}\
  \bibnamefont {Schuller}},\ and\ \bibinfo {author} {\bibfnamefont {J.~L.}\
  \bibnamefont {Vicent}},\ }\href {https://doi.org/10.1021/acsami.2c16950}
  {\bibfield  {journal} {\bibinfo  {journal} {ACS Applied Materials \&
  Interfaces}\ }\textbf {\bibinfo {volume} {14}},\ \bibinfo {pages} {54961}
  (\bibinfo {year} {2022})},\ \bibinfo {note} {pMID: 36469495},\ \Eprint
  {https://arxiv.org/abs/https://doi.org/10.1021/acsami.2c16950}
  {https://doi.org/10.1021/acsami.2c16950} \BibitemShut {NoStop}%
\bibitem [{\citenamefont {Kazakova}\ \emph {et~al.}(2019)\citenamefont
  {Kazakova}, \citenamefont {Puttock}, \citenamefont {Barton}, \citenamefont
  {Corte-León}, \citenamefont {Jaafar}, \citenamefont {Neu},\ and\
  \citenamefont {Asenjo}}]{2019_Frontiers}%
  \BibitemOpen
  \bibfield  {author} {\bibinfo {author} {\bibfnamefont {O.}~\bibnamefont
  {Kazakova}}, \bibinfo {author} {\bibfnamefont {R.}~\bibnamefont {Puttock}},
  \bibinfo {author} {\bibfnamefont {C.}~\bibnamefont {Barton}}, \bibinfo
  {author} {\bibfnamefont {H.}~\bibnamefont {Corte-León}}, \bibinfo {author}
  {\bibfnamefont {M.}~\bibnamefont {Jaafar}}, \bibinfo {author} {\bibfnamefont
  {V.}~\bibnamefont {Neu}},\ and\ \bibinfo {author} {\bibfnamefont
  {A.}~\bibnamefont {Asenjo}},\ }\href {https://doi.org/10.1063/1.5050712}
  {\bibfield  {journal} {\bibinfo  {journal} {Journal of Applied Physics}\
  }\textbf {\bibinfo {volume} {125}},\ \bibinfo {pages} {060901} (\bibinfo
  {year} {2019})}\BibitemShut {NoStop}%
\bibitem [{\citenamefont {Holl}\ \emph {et~al.}(2020)\citenamefont {Holl},
  \citenamefont {Knol}, \citenamefont {Pratzer}, \citenamefont {Chico},
  \citenamefont {Fernandes}, \citenamefont {Lounis},\ and\ \citenamefont
  {Morgenstern}}]{2020_Probing_the_pinning}%
  \BibitemOpen
  \bibfield  {author} {\bibinfo {author} {\bibfnamefont {C.}~\bibnamefont
  {Holl}}, \bibinfo {author} {\bibfnamefont {M.}~\bibnamefont {Knol}}, \bibinfo
  {author} {\bibfnamefont {M.}~\bibnamefont {Pratzer}}, \bibinfo {author}
  {\bibfnamefont {J.}~\bibnamefont {Chico}}, \bibinfo {author} {\bibfnamefont
  {I.~L.}\ \bibnamefont {Fernandes}}, \bibinfo {author} {\bibfnamefont
  {S.}~\bibnamefont {Lounis}},\ and\ \bibinfo {author} {\bibfnamefont
  {M.}~\bibnamefont {Morgenstern}},\ }\href
  {https://doi.org/10.1038/s41467-020-16701-y} {\bibfield  {journal} {\bibinfo
  {journal} {Nature Communications}\ }\textbf {\bibinfo {volume} {11}},\
  \bibinfo {pages} {2833} (\bibinfo {year} {2020})}\BibitemShut {NoStop}%
\bibitem [{\citenamefont {Wu}\ \emph {et~al.}(2010)\citenamefont {Wu},
  \citenamefont {Carlton}, \citenamefont {Oelker}, \citenamefont {Park},
  \citenamefont {Jin}, \citenamefont {Arenholz}, \citenamefont {Scholl},
  \citenamefont {Hwang}, \citenamefont {Bokor},\ and\ \citenamefont
  {Qiu}}]{2010_Switching_a_magnetic_vortex}%
  \BibitemOpen
  \bibfield  {author} {\bibinfo {author} {\bibfnamefont {J.}~\bibnamefont
  {Wu}}, \bibinfo {author} {\bibfnamefont {D.}~\bibnamefont {Carlton}},
  \bibinfo {author} {\bibfnamefont {E.}~\bibnamefont {Oelker}}, \bibinfo
  {author} {\bibfnamefont {J.~S.}\ \bibnamefont {Park}}, \bibinfo {author}
  {\bibfnamefont {E.}~\bibnamefont {Jin}}, \bibinfo {author} {\bibfnamefont
  {E.}~\bibnamefont {Arenholz}}, \bibinfo {author} {\bibfnamefont
  {A.}~\bibnamefont {Scholl}}, \bibinfo {author} {\bibfnamefont
  {C.}~\bibnamefont {Hwang}}, \bibinfo {author} {\bibfnamefont
  {J.}~\bibnamefont {Bokor}},\ and\ \bibinfo {author} {\bibfnamefont {Z.~Q.}\
  \bibnamefont {Qiu}},\ }\href {https://doi.org/10.1088/0953-8984/22/34/342001}
  {\bibfield  {journal} {\bibinfo  {journal} {Journal of Physics: Condensed
  Matter}\ }\textbf {\bibinfo {volume} {22}},\ \bibinfo {pages} {342001}
  (\bibinfo {year} {2010})}\BibitemShut {NoStop}%
\bibitem [{\citenamefont {Schneider}\ \emph {et~al.}(2000)\citenamefont
  {Schneider}, \citenamefont {Hoffmann},\ and\ \citenamefont
  {Zweck}}]{2000_Lorentz_microscopy}%
  \BibitemOpen
  \bibfield  {author} {\bibinfo {author} {\bibfnamefont {M.}~\bibnamefont
  {Schneider}}, \bibinfo {author} {\bibfnamefont {H.}~\bibnamefont
  {Hoffmann}},\ and\ \bibinfo {author} {\bibfnamefont {J.}~\bibnamefont
  {Zweck}},\ }\href {https://doi.org/10.1063/1.1320465} {\bibfield  {journal}
  {\bibinfo  {journal} {Applied Physics Letters}\ }\textbf {\bibinfo {volume}
  {77}},\ \bibinfo {pages} {2909} (\bibinfo {year} {2000})}\BibitemShut
  {NoStop}%
\bibitem [{\citenamefont {Geoffroy}\ \emph {et~al.}(1993)\citenamefont
  {Geoffroy}, \citenamefont {Givord}, \citenamefont {Otani}, \citenamefont
  {Pannetier}, \citenamefont {Santos}, \citenamefont {Schlenker},\ and\
  \citenamefont {Souche}}]{1993_TMOKE_hysteresis}%
  \BibitemOpen
  \bibfield  {author} {\bibinfo {author} {\bibfnamefont {O.}~\bibnamefont
  {Geoffroy}}, \bibinfo {author} {\bibfnamefont {D.}~\bibnamefont {Givord}},
  \bibinfo {author} {\bibfnamefont {Y.}~\bibnamefont {Otani}}, \bibinfo
  {author} {\bibfnamefont {B.}~\bibnamefont {Pannetier}}, \bibinfo {author}
  {\bibfnamefont {A.}~\bibnamefont {Santos}}, \bibinfo {author} {\bibfnamefont
  {M.}~\bibnamefont {Schlenker}},\ and\ \bibinfo {author} {\bibfnamefont
  {Y.}~\bibnamefont {Souche}},\ }\href
  {https://doi.org/https://doi.org/10.1016/0304-8853(93)91258-9} {\bibfield
  {journal} {\bibinfo  {journal} {Journal of Magnetism and Magnetic Materials}\
  }\textbf {\bibinfo {volume} {121}},\ \bibinfo {pages} {516} (\bibinfo {year}
  {1993})},\ \bibinfo {note} {proceedings of the International Symposium on
  Magnetic Ultrathin Films, Multilayers and Surfaces}\BibitemShut {NoStop}%
\bibitem [{\citenamefont {Costa-Krämer}\ \emph {et~al.}(2003)\citenamefont
  {Costa-Krämer}, \citenamefont {Guerrero}, \citenamefont {Melle},
  \citenamefont {García-Mochales},\ and\ \citenamefont
  {Briones}}]{2002_Pure_magneto_optic}%
  \BibitemOpen
  \bibfield  {author} {\bibinfo {author} {\bibfnamefont {J.~L.}\ \bibnamefont
  {Costa-Krämer}}, \bibinfo {author} {\bibfnamefont {C.}~\bibnamefont
  {Guerrero}}, \bibinfo {author} {\bibfnamefont {S.}~\bibnamefont {Melle}},
  \bibinfo {author} {\bibfnamefont {P.}~\bibnamefont {García-Mochales}},\ and\
  \bibinfo {author} {\bibfnamefont {F.}~\bibnamefont {Briones}},\ }\href
  {https://doi.org/10.1088/0957-4484/14/2/326} {\bibfield  {journal} {\bibinfo
  {journal} {Nanotechnology}\ }\textbf {\bibinfo {volume} {14}},\ \bibinfo
  {pages} {239} (\bibinfo {year} {2003})}\BibitemShut {NoStop}%
\bibitem [{\citenamefont {Grimsditch}\ and\ \citenamefont
  {Vavassori}(2004)}]{2004_The_diffracted_magnetoptic_Kerr}%
  \BibitemOpen
  \bibfield  {author} {\bibinfo {author} {\bibfnamefont {M.}~\bibnamefont
  {Grimsditch}}\ and\ \bibinfo {author} {\bibfnamefont {P.}~\bibnamefont
  {Vavassori}},\ }\href {https://doi.org/10.1088/0953-8984/16/9/R01} {\bibfield
   {journal} {\bibinfo  {journal} {Journal of Physics: Condensed Matter}\
  }\textbf {\bibinfo {volume} {16}},\ \bibinfo {pages} {R275} (\bibinfo {year}
  {2004})}\BibitemShut {NoStop}%
\bibitem [{\citenamefont {Westphalen}\ \emph {et~al.}(2007)\citenamefont
  {Westphalen}, \citenamefont {Lee}, \citenamefont {Remhof},\ and\
  \citenamefont {Zabel}}]{2007_Vector_and_Bragg_MOKE}%
  \BibitemOpen
  \bibfield  {author} {\bibinfo {author} {\bibfnamefont {A.}~\bibnamefont
  {Westphalen}}, \bibinfo {author} {\bibfnamefont {M.-S.}\ \bibnamefont {Lee}},
  \bibinfo {author} {\bibfnamefont {A.}~\bibnamefont {Remhof}},\ and\ \bibinfo
  {author} {\bibfnamefont {H.}~\bibnamefont {Zabel}},\ }\href
  {https://doi.org/10.1063/1.2821148} {\bibfield  {journal} {\bibinfo
  {journal} {Review of Scientific Instruments}\ }\textbf {\bibinfo {volume}
  {78}},\ \bibinfo {pages} {121301} (\bibinfo {year} {2007})},\ \Eprint
  {https://arxiv.org/abs/https://doi.org/10.1063/1.2821148}
  {https://doi.org/10.1063/1.2821148} \BibitemShut {NoStop}%
\bibitem [{\citenamefont {Lee}\ \emph {et~al.}(2008)\citenamefont {Lee},
  \citenamefont {Westphalen}, \citenamefont {Remhof}, \citenamefont
  {Schumann},\ and\ \citenamefont {Zabel}}]{2008_Extended_longitudinal}%
  \BibitemOpen
  \bibfield  {author} {\bibinfo {author} {\bibfnamefont {M.-S.}\ \bibnamefont
  {Lee}}, \bibinfo {author} {\bibfnamefont {A.}~\bibnamefont {Westphalen}},
  \bibinfo {author} {\bibfnamefont {A.}~\bibnamefont {Remhof}}, \bibinfo
  {author} {\bibfnamefont {A.}~\bibnamefont {Schumann}},\ and\ \bibinfo
  {author} {\bibfnamefont {H.}~\bibnamefont {Zabel}},\ }\href
  {https://doi.org/10.1063/1.2919160} {\bibfield  {journal} {\bibinfo
  {journal} {Journal of Applied Physics}\ }\textbf {\bibinfo {volume} {103}},\
  \bibinfo {pages} {093913} (\bibinfo {year} {2008})},\ \Eprint
  {https://arxiv.org/abs/https://doi.org/10.1063/1.2919160}
  {https://doi.org/10.1063/1.2919160} \BibitemShut {NoStop}%
\bibitem [{\citenamefont {{van der
  Laan}}(2008)}]{2008_Soft_X_ray_resonant_magnetic}%
  \BibitemOpen
  \bibfield  {author} {\bibinfo {author} {\bibfnamefont {G.}~\bibnamefont {{van
  der Laan}}},\ }\href
  {https://doi.org/https://doi.org/10.1016/j.crhy.2007.06.004} {\bibfield
  {journal} {\bibinfo  {journal} {Comptes Rendus Physique}\ }\textbf {\bibinfo
  {volume} {9}},\ \bibinfo {pages} {570} (\bibinfo {year} {2008})},\ \bibinfo
  {note} {synchrotron x-rays and condensed matter}\BibitemShut {NoStop}%
\bibitem [{\citenamefont {Choi}\ \emph {et~al.}(2006)\citenamefont {Choi},
  \citenamefont {Lee}, \citenamefont {Freeland}, \citenamefont {Srajer},\ and\
  \citenamefont {Metlushko}}]{2006_Layer_resolved}%
  \BibitemOpen
  \bibfield  {author} {\bibinfo {author} {\bibfnamefont {Y.}~\bibnamefont
  {Choi}}, \bibinfo {author} {\bibfnamefont {D.~R.}\ \bibnamefont {Lee}},
  \bibinfo {author} {\bibfnamefont {J.~W.}\ \bibnamefont {Freeland}}, \bibinfo
  {author} {\bibfnamefont {G.}~\bibnamefont {Srajer}},\ and\ \bibinfo {author}
  {\bibfnamefont {V.}~\bibnamefont {Metlushko}},\ }\href
  {https://doi.org/10.1063/1.2179116} {\bibfield  {journal} {\bibinfo
  {journal} {Applied Physics Letters}\ }\textbf {\bibinfo {volume} {88}},\
  \bibinfo {pages} {112502} (\bibinfo {year} {2006})},\ \Eprint
  {https://arxiv.org/abs/https://doi.org/10.1063/1.2179116}
  {https://doi.org/10.1063/1.2179116} \BibitemShut {NoStop}%
\bibitem [{\citenamefont {Rose}\ \emph {et~al.}(2007)\citenamefont {Rose},
  \citenamefont {Cheng}, \citenamefont {Keavney}, \citenamefont {Freeland},
  \citenamefont {Buchanan}, \citenamefont {Ilic},\ and\ \citenamefont
  {Metlushko}}]{2007_The_breakdown}%
  \BibitemOpen
  \bibfield  {author} {\bibinfo {author} {\bibfnamefont {V.}~\bibnamefont
  {Rose}}, \bibinfo {author} {\bibfnamefont {X.~M.}\ \bibnamefont {Cheng}},
  \bibinfo {author} {\bibfnamefont {D.~J.}\ \bibnamefont {Keavney}}, \bibinfo
  {author} {\bibfnamefont {J.~W.}\ \bibnamefont {Freeland}}, \bibinfo {author}
  {\bibfnamefont {K.~S.}\ \bibnamefont {Buchanan}}, \bibinfo {author}
  {\bibfnamefont {B.}~\bibnamefont {Ilic}},\ and\ \bibinfo {author}
  {\bibfnamefont {V.}~\bibnamefont {Metlushko}},\ }\href
  {https://doi.org/10.1063/1.2786856} {\bibfield  {journal} {\bibinfo
  {journal} {Applied Physics Letters}\ }\textbf {\bibinfo {volume} {91}},\
  \bibinfo {pages} {132501} (\bibinfo {year} {2007})},\ \Eprint
  {https://arxiv.org/abs/https://doi.org/10.1063/1.2786856}
  {https://doi.org/10.1063/1.2786856} \BibitemShut {NoStop}%
\bibitem [{\citenamefont {Lee}\ \emph {et~al.}(2007)\citenamefont {Lee},
  \citenamefont {Freeland}, \citenamefont {Choi}, \citenamefont {Srajer},
  \citenamefont {Metlushko},\ and\ \citenamefont
  {Ilic}}]{2007_X_ray_resonant_magnetic_scattering}%
  \BibitemOpen
  \bibfield  {author} {\bibinfo {author} {\bibfnamefont {D.~R.}\ \bibnamefont
  {Lee}}, \bibinfo {author} {\bibfnamefont {J.~W.}\ \bibnamefont {Freeland}},
  \bibinfo {author} {\bibfnamefont {Y.}~\bibnamefont {Choi}}, \bibinfo {author}
  {\bibfnamefont {G.}~\bibnamefont {Srajer}}, \bibinfo {author} {\bibfnamefont
  {V.}~\bibnamefont {Metlushko}},\ and\ \bibinfo {author} {\bibfnamefont
  {B.}~\bibnamefont {Ilic}},\ }\href
  {https://doi.org/10.1103/PhysRevB.76.144425} {\bibfield  {journal} {\bibinfo
  {journal} {Phys. Rev. B}\ }\textbf {\bibinfo {volume} {76}},\ \bibinfo
  {pages} {144425} (\bibinfo {year} {2007})}\BibitemShut {NoStop}%
\bibitem [{\citenamefont {Díaz}\ \emph {et~al.}(2019)\citenamefont {Díaz},
  \citenamefont {Gargiani}, \citenamefont {Quirós}, \citenamefont {Redondo},
  \citenamefont {Morales}, \citenamefont {Álvarez Prado}, \citenamefont
  {Martín}, \citenamefont {Scholl}, \citenamefont {Ferrer}, \citenamefont
  {Vélez},\ and\ \citenamefont {Valvidares}}]{chiral_nanotech}%
  \BibitemOpen
  \bibfield  {author} {\bibinfo {author} {\bibfnamefont {J.}~\bibnamefont
  {Díaz}}, \bibinfo {author} {\bibfnamefont {P.}~\bibnamefont {Gargiani}},
  \bibinfo {author} {\bibfnamefont {C.}~\bibnamefont {Quirós}}, \bibinfo
  {author} {\bibfnamefont {C.}~\bibnamefont {Redondo}}, \bibinfo {author}
  {\bibfnamefont {R.}~\bibnamefont {Morales}}, \bibinfo {author} {\bibfnamefont
  {L.~M.}\ \bibnamefont {Álvarez Prado}}, \bibinfo {author} {\bibfnamefont
  {J.~I.}\ \bibnamefont {Martín}}, \bibinfo {author} {\bibfnamefont
  {A.}~\bibnamefont {Scholl}}, \bibinfo {author} {\bibfnamefont
  {S.}~\bibnamefont {Ferrer}}, \bibinfo {author} {\bibfnamefont
  {M.}~\bibnamefont {Vélez}},\ and\ \bibinfo {author} {\bibfnamefont {S.~M.}\
  \bibnamefont {Valvidares}},\ }\href
  {https://doi.org/10.1088/1361-6528/ab46d7} {\bibfield  {journal} {\bibinfo
  {journal} {Nanotechnology}\ }\textbf {\bibinfo {volume} {31}},\ \bibinfo
  {pages} {025702} (\bibinfo {year} {2019})}\BibitemShut {NoStop}%
\bibitem [{\citenamefont {Mora}\ \emph {et~al.}(2018)\citenamefont {Mora},
  \citenamefont {Perez-Valle}, \citenamefont {Redondo}, \citenamefont
  {Boyano},\ and\ \citenamefont {Morales}}]{2018_Cost_Effective}%
  \BibitemOpen
  \bibfield  {author} {\bibinfo {author} {\bibfnamefont {B.}~\bibnamefont
  {Mora}}, \bibinfo {author} {\bibfnamefont {A.}~\bibnamefont {Perez-Valle}},
  \bibinfo {author} {\bibfnamefont {C.}~\bibnamefont {Redondo}}, \bibinfo
  {author} {\bibfnamefont {M.~D.}\ \bibnamefont {Boyano}},\ and\ \bibinfo
  {author} {\bibfnamefont {R.}~\bibnamefont {Morales}},\ }\href
  {https://doi.org/10.1021/acsami.7b16779} {\bibfield  {journal} {\bibinfo
  {journal} {ACS applied materials and amp; interfaces}\ }\textbf {\bibinfo
  {volume} {10}},\ \bibinfo {pages} {8165—8172} (\bibinfo {year}
  {2018})}\BibitemShut {NoStop}%
\bibitem [{\citenamefont {Elzo}\ \emph {et~al.}(2012)\citenamefont {Elzo},
  \citenamefont {Jal}, \citenamefont {Bunau}, \citenamefont {Grenier},
  \citenamefont {Joly}, \citenamefont {Ramos}, \citenamefont {Tolentino},
  \citenamefont {Tonnerre},\ and\ \citenamefont
  {Jaouen}}]{2011_X_ray_resonant_magnetic}%
  \BibitemOpen
  \bibfield  {author} {\bibinfo {author} {\bibfnamefont {M.}~\bibnamefont
  {Elzo}}, \bibinfo {author} {\bibfnamefont {E.}~\bibnamefont {Jal}}, \bibinfo
  {author} {\bibfnamefont {O.}~\bibnamefont {Bunau}}, \bibinfo {author}
  {\bibfnamefont {S.}~\bibnamefont {Grenier}}, \bibinfo {author} {\bibfnamefont
  {Y.}~\bibnamefont {Joly}}, \bibinfo {author} {\bibfnamefont {A.}~\bibnamefont
  {Ramos}}, \bibinfo {author} {\bibfnamefont {H.}~\bibnamefont {Tolentino}},
  \bibinfo {author} {\bibfnamefont {J.}~\bibnamefont {Tonnerre}},\ and\
  \bibinfo {author} {\bibfnamefont {N.}~\bibnamefont {Jaouen}},\ }\href
  {https://doi.org/https://doi.org/10.1016/j.jmmm.2011.07.019} {\bibfield
  {journal} {\bibinfo  {journal} {Journal of Magnetism and Magnetic Materials}\
  }\textbf {\bibinfo {volume} {324}},\ \bibinfo {pages} {105} (\bibinfo {year}
  {2012})}\BibitemShut {NoStop}%
\bibitem [{\citenamefont {Lazić}\ \emph {et~al.}(2013)\citenamefont {Lazić},
  \citenamefont {Chamritski}, \citenamefont {Pooke}, \citenamefont
  {Valvidares}, \citenamefont {Pellegrin}, \citenamefont {Ferrer},
  \citenamefont {Granados},\ and\ \citenamefont {Obradors}}]{BOREAS_setup}%
  \BibitemOpen
  \bibfield  {author} {\bibinfo {author} {\bibfnamefont {Z.}~\bibnamefont
  {Lazić}}, \bibinfo {author} {\bibfnamefont {V.}~\bibnamefont {Chamritski}},
  \bibinfo {author} {\bibfnamefont {D.}~\bibnamefont {Pooke}}, \bibinfo
  {author} {\bibfnamefont {S.~M.}\ \bibnamefont {Valvidares}}, \bibinfo
  {author} {\bibfnamefont {E.}~\bibnamefont {Pellegrin}}, \bibinfo {author}
  {\bibfnamefont {S.}~\bibnamefont {Ferrer}}, \bibinfo {author} {\bibfnamefont
  {X.}~\bibnamefont {Granados}},\ and\ \bibinfo {author} {\bibfnamefont
  {X.}~\bibnamefont {Obradors}},\ }\href
  {https://doi.org/10.1088/1742-6596/425/10/102003} {\bibfield  {journal}
  {\bibinfo  {journal} {Journal of Physics: Conference Series}\ }\textbf
  {\bibinfo {volume} {425}},\ \bibinfo {pages} {102003} (\bibinfo {year}
  {2013})}\BibitemShut {NoStop}%
\bibitem [{\citenamefont {Lee}\ \emph {et~al.}(2003)\citenamefont {Lee},
  \citenamefont {Sinha}, \citenamefont {Haskel}, \citenamefont {Choi},
  \citenamefont {Lang}, \citenamefont {Stepanov},\ and\ \citenamefont
  {Srajer}}]{2003_XRMS_from_structurally_and_magnetically_rough_interfaces}%
  \BibitemOpen
  \bibfield  {author} {\bibinfo {author} {\bibfnamefont {D.~R.}\ \bibnamefont
  {Lee}}, \bibinfo {author} {\bibfnamefont {S.~K.}\ \bibnamefont {Sinha}},
  \bibinfo {author} {\bibfnamefont {D.}~\bibnamefont {Haskel}}, \bibinfo
  {author} {\bibfnamefont {Y.}~\bibnamefont {Choi}}, \bibinfo {author}
  {\bibfnamefont {J.~C.}\ \bibnamefont {Lang}}, \bibinfo {author}
  {\bibfnamefont {S.~A.}\ \bibnamefont {Stepanov}},\ and\ \bibinfo {author}
  {\bibfnamefont {G.}~\bibnamefont {Srajer}},\ }\href
  {https://doi.org/10.1103/PhysRevB.68.224409} {\bibfield  {journal} {\bibinfo
  {journal} {Phys. Rev. B}\ }\textbf {\bibinfo {volume} {68}},\ \bibinfo
  {pages} {224409} (\bibinfo {year} {2003})}\BibitemShut {NoStop}%
\bibitem [{\citenamefont {Schrag}\ \emph {et~al.}(2000)\citenamefont {Schrag},
  \citenamefont {Anguelouch}, \citenamefont {Ingvarsson}, \citenamefont {Xiao},
  \citenamefont {Lu}, \citenamefont {Trouilloud}, \citenamefont {Gupta},
  \citenamefont {Wanner}, \citenamefont {Gallagher}, \citenamefont {Rice},\
  and\ \citenamefont {Parkin}}]{2000_Neel_orange_peel_coupling}%
  \BibitemOpen
  \bibfield  {author} {\bibinfo {author} {\bibfnamefont {B.~D.}\ \bibnamefont
  {Schrag}}, \bibinfo {author} {\bibfnamefont {A.}~\bibnamefont {Anguelouch}},
  \bibinfo {author} {\bibfnamefont {S.}~\bibnamefont {Ingvarsson}}, \bibinfo
  {author} {\bibfnamefont {G.}~\bibnamefont {Xiao}}, \bibinfo {author}
  {\bibfnamefont {Y.}~\bibnamefont {Lu}}, \bibinfo {author} {\bibfnamefont
  {P.~L.}\ \bibnamefont {Trouilloud}}, \bibinfo {author} {\bibfnamefont
  {A.}~\bibnamefont {Gupta}}, \bibinfo {author} {\bibfnamefont {R.~A.}\
  \bibnamefont {Wanner}}, \bibinfo {author} {\bibfnamefont {W.~J.}\
  \bibnamefont {Gallagher}}, \bibinfo {author} {\bibfnamefont {P.~M.}\
  \bibnamefont {Rice}},\ and\ \bibinfo {author} {\bibfnamefont {S.~S.~P.}\
  \bibnamefont {Parkin}},\ }\href {https://doi.org/10.1063/1.1315633}
  {\bibfield  {journal} {\bibinfo  {journal} {Applied Physics Letters}\
  }\textbf {\bibinfo {volume} {77}},\ \bibinfo {pages} {2373} (\bibinfo {year}
  {2000})},\ \Eprint {https://arxiv.org/abs/https://doi.org/10.1063/1.1315633}
  {https://doi.org/10.1063/1.1315633} \BibitemShut {NoStop}%
\bibitem [{\citenamefont {Dumas}\ \emph {et~al.}(2011)\citenamefont {Dumas},
  \citenamefont {Gilbert}, \citenamefont {Eibagi},\ and\ \citenamefont
  {Liu}}]{2011_Chirality_control}%
  \BibitemOpen
  \bibfield  {author} {\bibinfo {author} {\bibfnamefont {R.~K.}\ \bibnamefont
  {Dumas}}, \bibinfo {author} {\bibfnamefont {D.~A.}\ \bibnamefont {Gilbert}},
  \bibinfo {author} {\bibfnamefont {N.}~\bibnamefont {Eibagi}},\ and\ \bibinfo
  {author} {\bibfnamefont {K.}~\bibnamefont {Liu}},\ }\href
  {https://doi.org/10.1103/PhysRevB.83.060415} {\bibfield  {journal} {\bibinfo
  {journal} {Phys. Rev. B}\ }\textbf {\bibinfo {volume} {83}},\ \bibinfo
  {pages} {060415} (\bibinfo {year} {2011})}\BibitemShut {NoStop}%
\bibitem [{\citenamefont {Jaafar}\ \emph {et~al.}(2010)\citenamefont {Jaafar},
  \citenamefont {Yanes}, \citenamefont {Perez~de Lara}, \citenamefont
  {Chubykalo-Fesenko}, \citenamefont {Asenjo}, \citenamefont {Gonzalez},
  \citenamefont {Anguita}, \citenamefont {Vazquez},\ and\ \citenamefont
  {Vicent}}]{2010_Control_of_the_chirality}%
  \BibitemOpen
  \bibfield  {author} {\bibinfo {author} {\bibfnamefont {M.}~\bibnamefont
  {Jaafar}}, \bibinfo {author} {\bibfnamefont {R.}~\bibnamefont {Yanes}},
  \bibinfo {author} {\bibfnamefont {D.}~\bibnamefont {Perez~de Lara}}, \bibinfo
  {author} {\bibfnamefont {O.}~\bibnamefont {Chubykalo-Fesenko}}, \bibinfo
  {author} {\bibfnamefont {A.}~\bibnamefont {Asenjo}}, \bibinfo {author}
  {\bibfnamefont {E.~M.}\ \bibnamefont {Gonzalez}}, \bibinfo {author}
  {\bibfnamefont {J.~V.}\ \bibnamefont {Anguita}}, \bibinfo {author}
  {\bibfnamefont {M.}~\bibnamefont {Vazquez}},\ and\ \bibinfo {author}
  {\bibfnamefont {J.~L.}\ \bibnamefont {Vicent}},\ }\href
  {https://doi.org/10.1103/PhysRevB.81.054439} {\bibfield  {journal} {\bibinfo
  {journal} {Phys. Rev. B}\ }\textbf {\bibinfo {volume} {81}},\ \bibinfo
  {pages} {054439} (\bibinfo {year} {2010})}\BibitemShut {NoStop}%
\bibitem [{\citenamefont {Agramunt-Puig}\ \emph {et~al.}(2014)\citenamefont
  {Agramunt-Puig}, \citenamefont {Del-Valle}, \citenamefont {Navau},\ and\
  \citenamefont {Sanchez}}]{2014_Controlling_vortex_chirality}%
  \BibitemOpen
  \bibfield  {author} {\bibinfo {author} {\bibfnamefont {S.}~\bibnamefont
  {Agramunt-Puig}}, \bibinfo {author} {\bibfnamefont {N.}~\bibnamefont
  {Del-Valle}}, \bibinfo {author} {\bibfnamefont {C.}~\bibnamefont {Navau}},\
  and\ \bibinfo {author} {\bibfnamefont {A.}~\bibnamefont {Sanchez}},\ }\href
  {https://doi.org/10.1063/1.4861423} {\bibfield  {journal} {\bibinfo
  {journal} {Applied Physics Letters}\ }\textbf {\bibinfo {volume} {104}},\
  \bibinfo {pages} {012407} (\bibinfo {year} {2014})},\ \Eprint
  {https://arxiv.org/abs/https://doi.org/10.1063/1.4861423}
  {https://doi.org/10.1063/1.4861423} \BibitemShut {NoStop}%
\bibitem [{\citenamefont {Cambel}\ and\ \citenamefont
  {Karapetrov}(2011)}]{2011_Control_of_vortex_chirality}%
  \BibitemOpen
  \bibfield  {author} {\bibinfo {author} {\bibfnamefont {V.}~\bibnamefont
  {Cambel}}\ and\ \bibinfo {author} {\bibfnamefont {G.}~\bibnamefont
  {Karapetrov}},\ }\href {https://doi.org/10.1103/PhysRevB.84.014424}
  {\bibfield  {journal} {\bibinfo  {journal} {Phys. Rev. B}\ }\textbf {\bibinfo
  {volume} {84}},\ \bibinfo {pages} {014424} (\bibinfo {year}
  {2011})}\BibitemShut {NoStop}%
\bibitem [{\citenamefont {Zhong}\ \emph {et~al.}(2009)\citenamefont {Zhong},
  \citenamefont {Zhang}, \citenamefont {Tang}, \citenamefont {Jing},
  \citenamefont {Jia},\ and\ \citenamefont
  {Liu}}]{2009_Vortex_chirality_control}%
  \BibitemOpen
  \bibfield  {author} {\bibinfo {author} {\bibfnamefont {Z.}~\bibnamefont
  {Zhong}}, \bibinfo {author} {\bibfnamefont {H.}~\bibnamefont {Zhang}},
  \bibinfo {author} {\bibfnamefont {X.}~\bibnamefont {Tang}}, \bibinfo {author}
  {\bibfnamefont {Y.}~\bibnamefont {Jing}}, \bibinfo {author} {\bibfnamefont
  {L.}~\bibnamefont {Jia}},\ and\ \bibinfo {author} {\bibfnamefont
  {S.}~\bibnamefont {Liu}},\ }\href
  {https://doi.org/https://doi.org/10.1016/j.jmmm.2009.02.030} {\bibfield
  {journal} {\bibinfo  {journal} {Journal of Magnetism and Magnetic Materials}\
  }\textbf {\bibinfo {volume} {321}},\ \bibinfo {pages} {2345} (\bibinfo {year}
  {2009})}\BibitemShut {NoStop}%
\end{thebibliography}

 %

\end{document}